\documentclass[sigconf]{acmart}

\copyrightyear{2021} 
\acmYear{2021} 
\setcopyright{acmcopyright}
\acmConference[KDD '21]{Proceedings of the 27th ACM SIGKDD Conference on Knowledge Discovery and Data Mining}{August 14--18, 2021}{Virtual Event, Singapore}
\acmBooktitle{Proceedings of the 27th ACM SIGKDD Conference on Knowledge Discovery and Data Mining (KDD '21), August 14--18, 2021, Virtual Event, Singapore}
\acmPrice{15.00}
\acmDOI{10.1145/3447548.3467279}
\acmISBN{978-1-4503-8332-5/21/08}

\usepackage{subfig}
\usepackage{booktabs} 
\usepackage{amsmath}
\usepackage{amsthm}
\usepackage{amsfonts}       
\usepackage{nicefrac}       
\usepackage{xspace}
\usepackage{xcolor}
\usepackage{backnaur}
\usepackage{multirow}
\usepackage{makecell}
\usepackage{appendix}
\usepackage{enumitem}
\usepackage{balance}

\theoremstyle{definition}
\newtheorem{definition}{Definition}

\newcommand{\reffig}[1]{Figure~\ref{#1}}
\newcommand{\refsec}[1]{\S\ref{#1}} 
\newcommand{\reftab}[1]{Table~\ref{#1}}

\def\eg{\textit{e.g.}\xspace}
\def\Eg{\textit{E.g.}\xspace}
\def\etal{\textit{et al.}\xspace}
\def\etc{\textit{etc.}\xspace}
\def\ie{\textit{i.e.}\xspace}

\def\wrt{\textit{w.r.t.}\xspace}

\begin{document}
\fancyhead{}

\title[Table2Charts: Recommending Charts by Learning Shared Table Representations]{Table2Charts: Recommending Charts by~Learning~Shared~Table~Representations}

\author{Mengyu Zhou}
\authornote{Author emails: \{mezho, jiwe, shihan, yinichen, djiang, dongmeiz\}@microsoft.com.}
\orcid{0000-0002-0322-7513}
\affiliation{%
  \institution{Microsoft Research}
}
\author{Qingtao Li}
\authornote{The contributions by Qingtao Li, Xinyi He, Yuejiang Li and Yibo Liu have been conducted and completed during their internships at Microsoft Research Asia, Beijing, China. Their school emails are: newdaylqt@pku.edu.cn, hxyhxy@stu.xjtu.edu.cn, lyj18@mails.tsinghua.edu.cn, and yl6769@nyu.edu.}
\affiliation{%
  \institution{Peking University}
}
\author{Xinyi He}
\authornotemark[2]
\affiliation{%
  \institution{Xi'an Jiaotong University}
}
\author{Yuejiang Li}
\authornotemark[2]
\affiliation{%
  \institution{Tsinghua University}
}
\author{Yibo Liu}
\authornotemark[2]
\affiliation{%
  \institution{New York University}
}
\author{Wei Ji}
\authornotemark[1]
\affiliation{%
  \institution{Microsoft}
}
\author{Shi Han}
\authornotemark[1]
\affiliation{%
  \institution{Microsoft Research}
  \city{Beijing}
  \country{China}
}
\author{Yining Chen}
\authornotemark[1]
\author{Daxin Jiang}
\authornotemark[1]
\affiliation{%
  \institution{Microsoft}
}
\author{Dongmei Zhang}
\authornotemark[1]
\affiliation{%
  \institution{Microsoft Research}
  \city{Beijing}
  \country{China}
}

\renewcommand{\shortauthors}{Mengyu Zhou \etal}

\begin{abstract}
It is common for people to create different types of charts to explore a multi-dimensional dataset (table). However, to recommend commonly composed charts in real world, one should take the challenges of efficiency, imbalanced data and table context into consideration.
In this paper, we propose Table2Charts framework\footnote{Code will be published at \url{https://github.com/microsoft/Table2Charts} to facilitate future research, once it is approved by an internal review.} which learns common patterns from a large corpus of (table, charts) pairs. Based on deep Q-learning with copying mechanism and heuristic searching, Table2Charts does table-to-sequence generation, where each sequence follows a chart template.
On a large spreadsheet corpus with 165k tables and 266k charts, we show that Table2Charts could learn a shared representation of table fields so that recommendation tasks on different chart types could mutually enhance each other. Table2Charts outperforms other chart recommendation systems in both multi-type task (with doubled recall numbers R@3=$0.61$ and R@1=$0.43$) and human evaluations.

\end{abstract}



\maketitle

\section{Introduction}
\label{sec:intro}

\begin{figure*}[tbp]
	\centering
	\subfloat[Bar Chart for \reftab{tab:A}. \label{fig:bar-example}]{\includegraphics[width=0.31\linewidth]{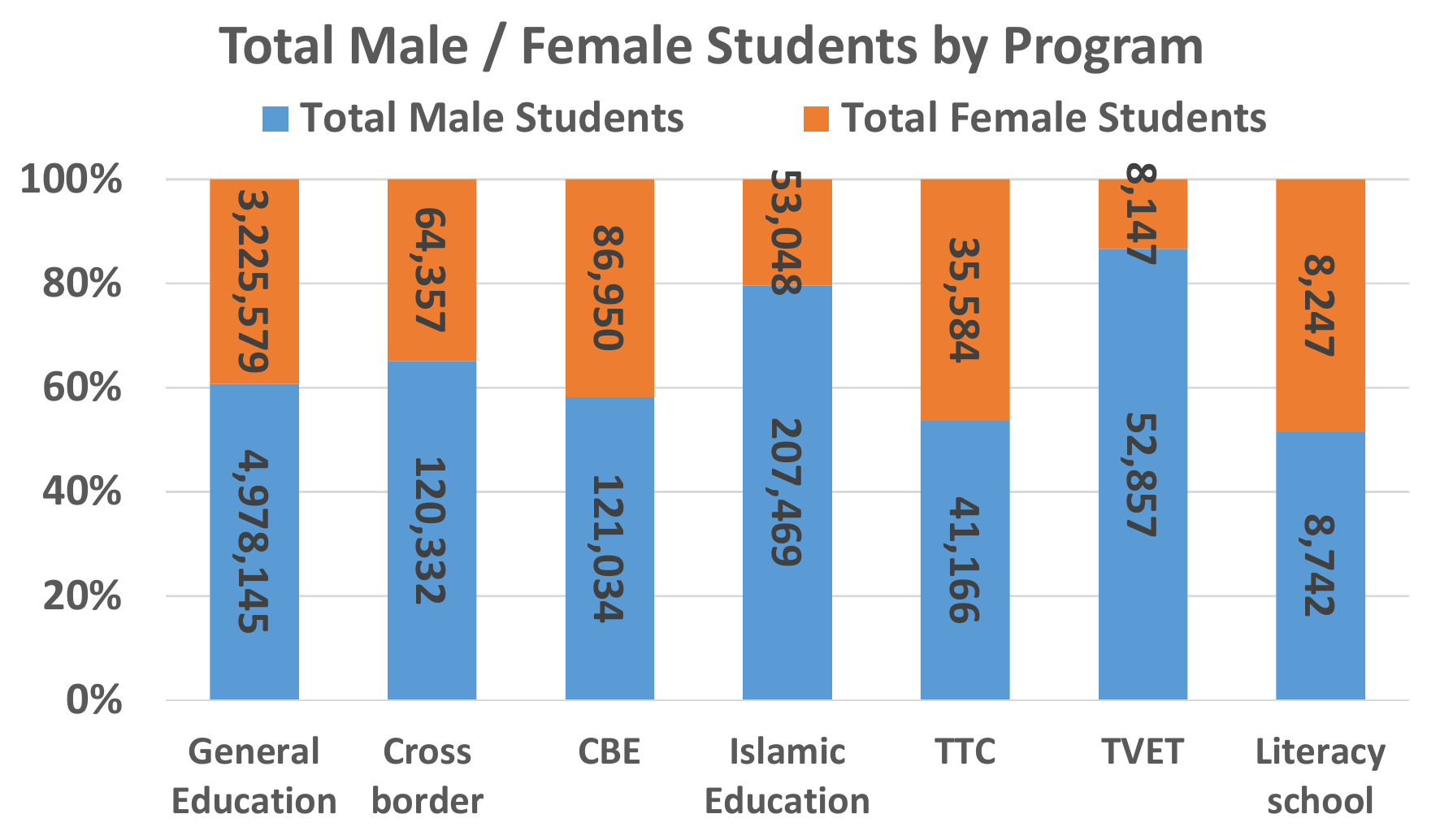}\label{fig:bar}}\quad
	\subfloat[Area Chart for \reftab{tab:A}.]{\includegraphics[width=0.31\linewidth]{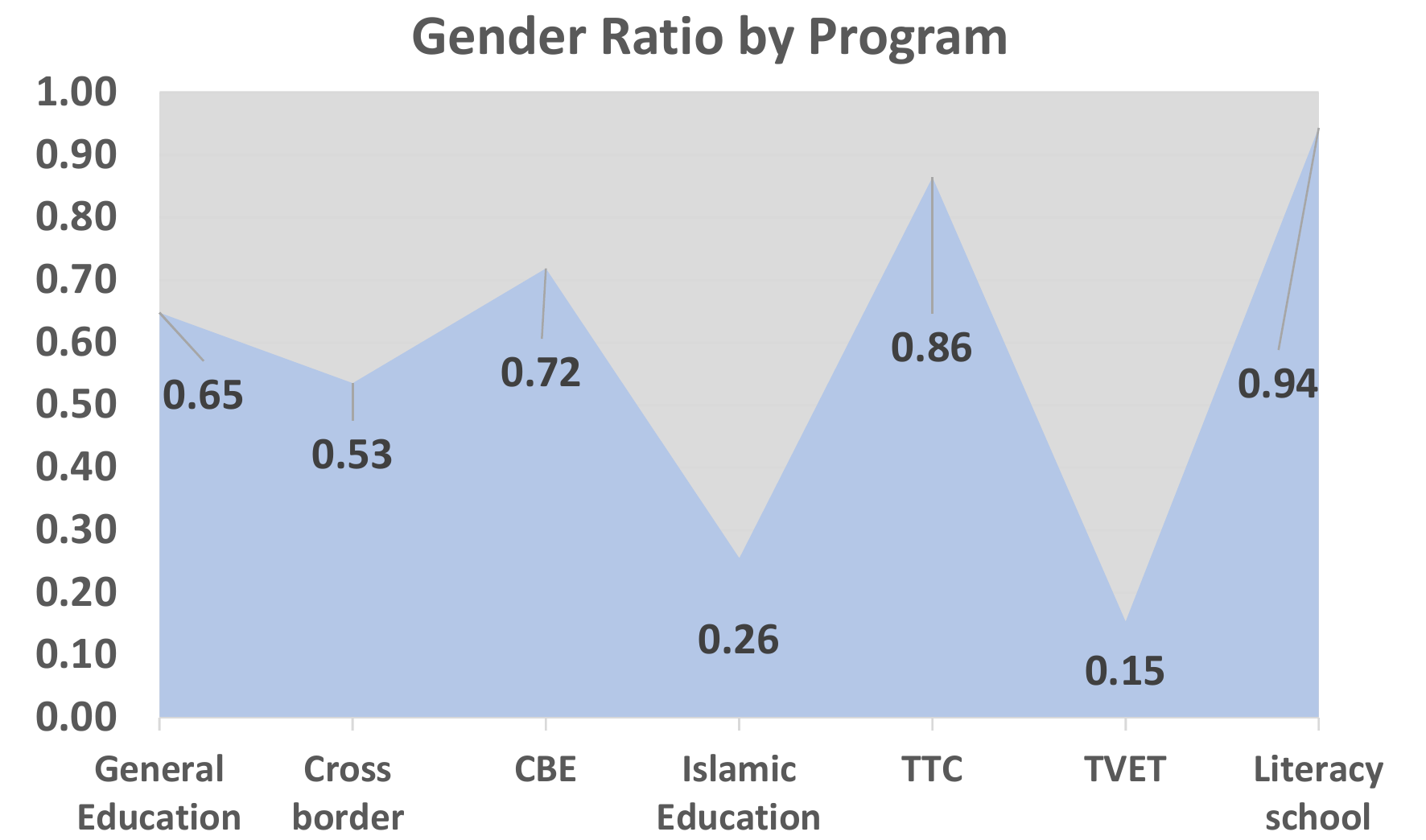}\label{fig:area}}\quad
	\subfloat[Scatter Chart for \reftab{tab:B}.]{\includegraphics[width=0.31\linewidth]{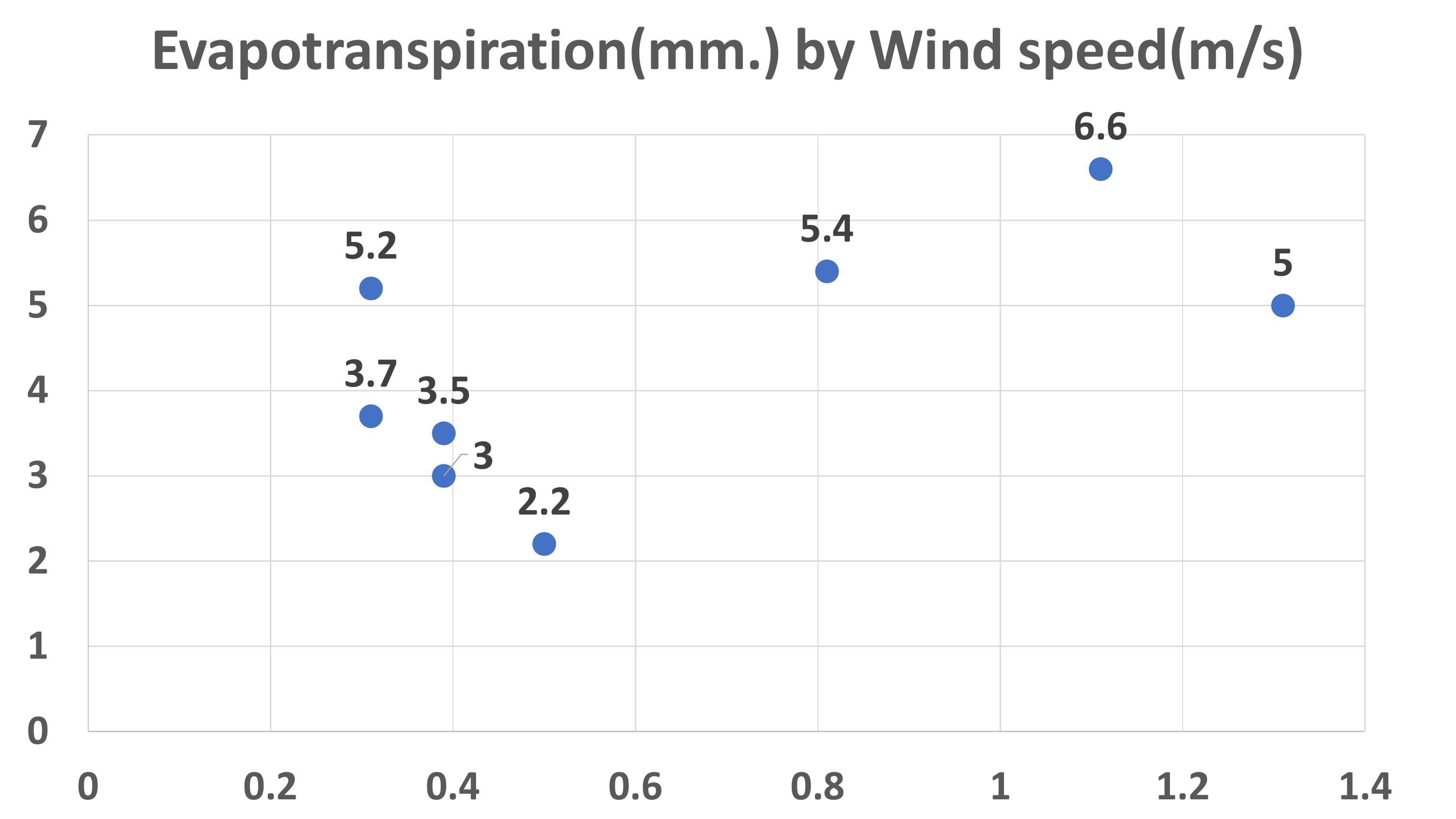}\label{fig:series}}\\
	\subfloat[Pie Chart for \reftab{tab:A}.]{\includegraphics[width=0.31\linewidth]{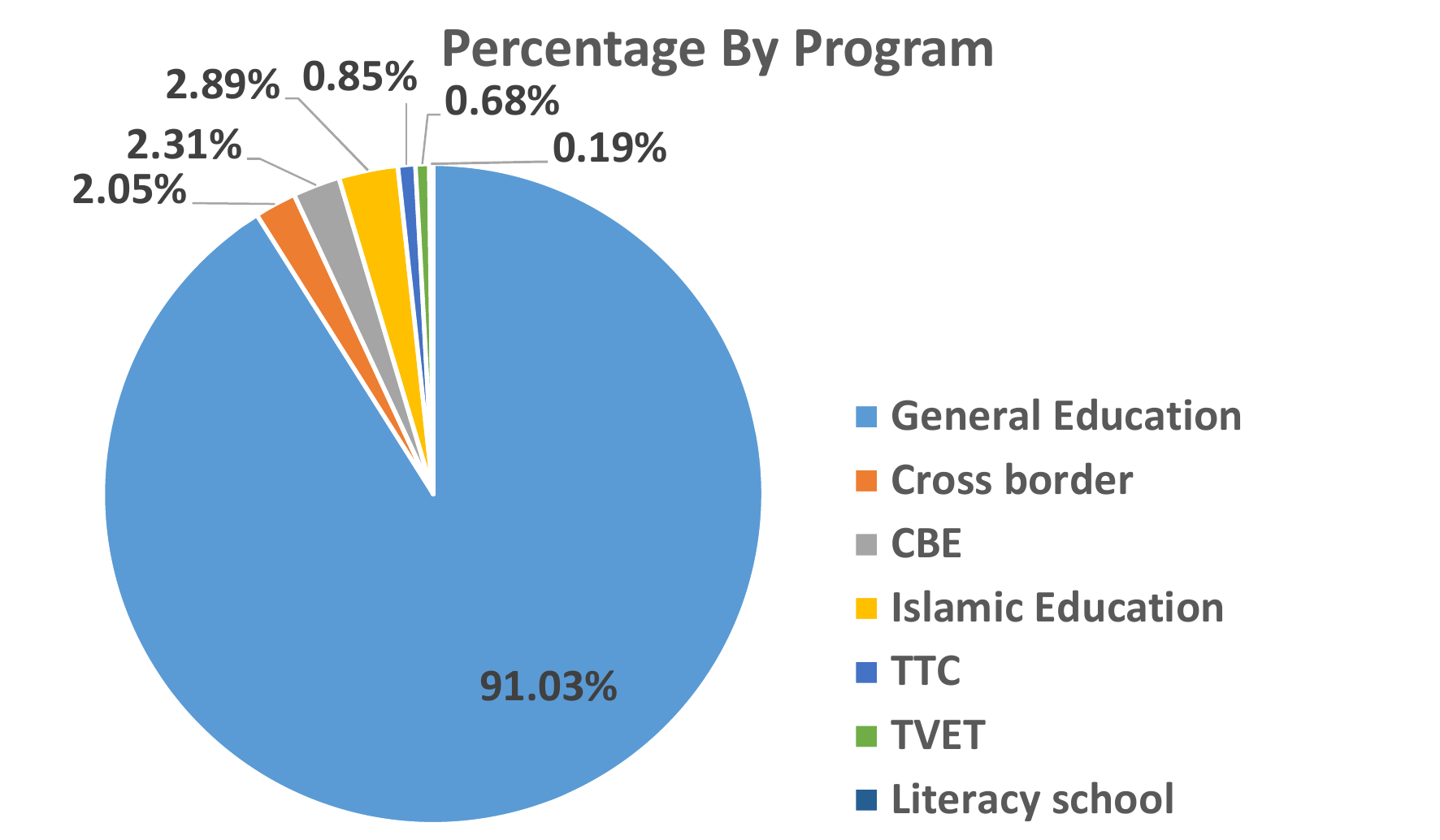}\label{fig:pie}}\quad
	\subfloat[Line Chart for \reftab{tab:B}. \label{fig:line-example}]{\includegraphics[width=0.31\linewidth]{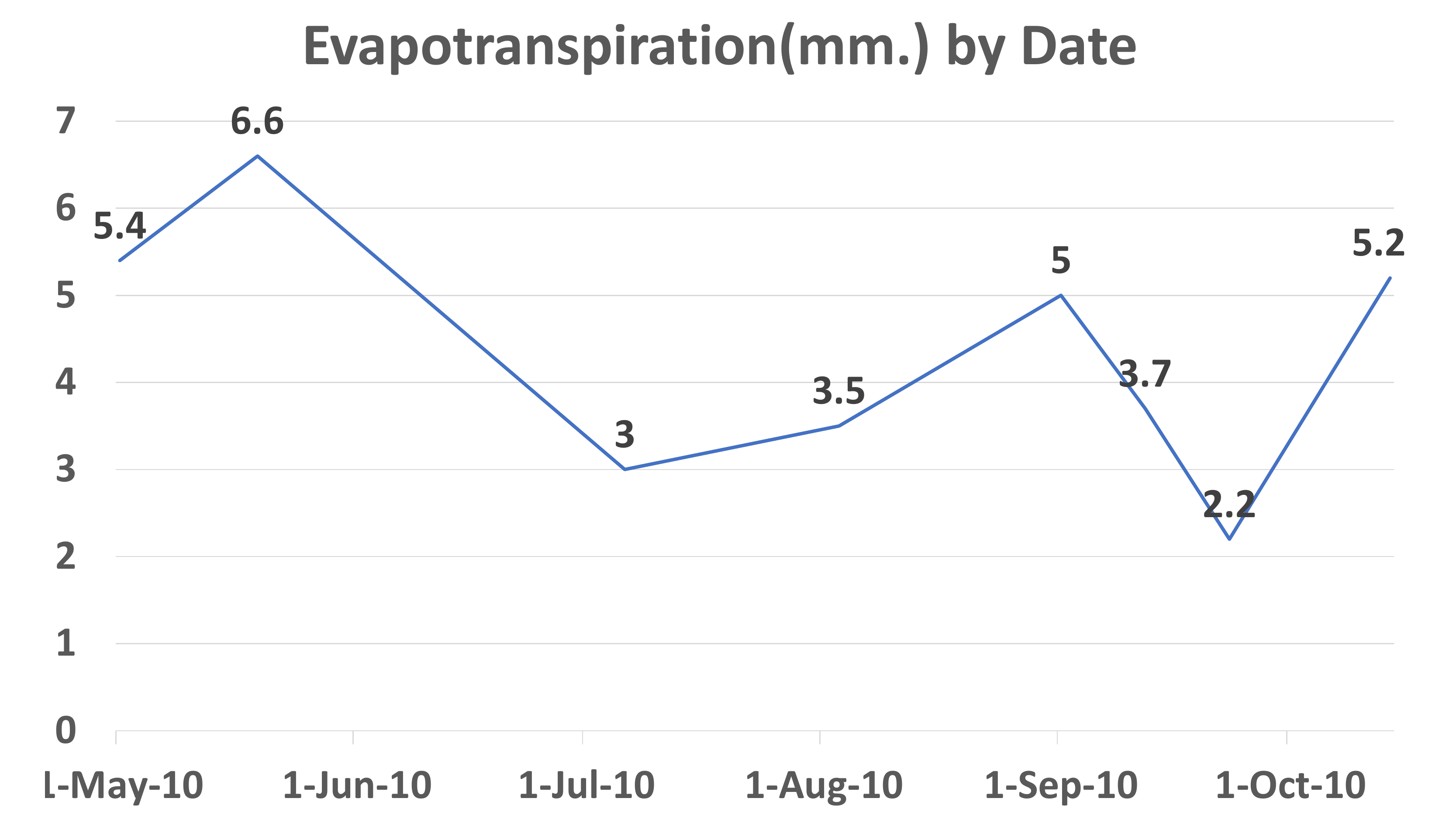}\label{fig:line}}\quad
	\subfloat[Radar Chart for \reftab{tab:B}.]{\includegraphics[width=0.31\linewidth]{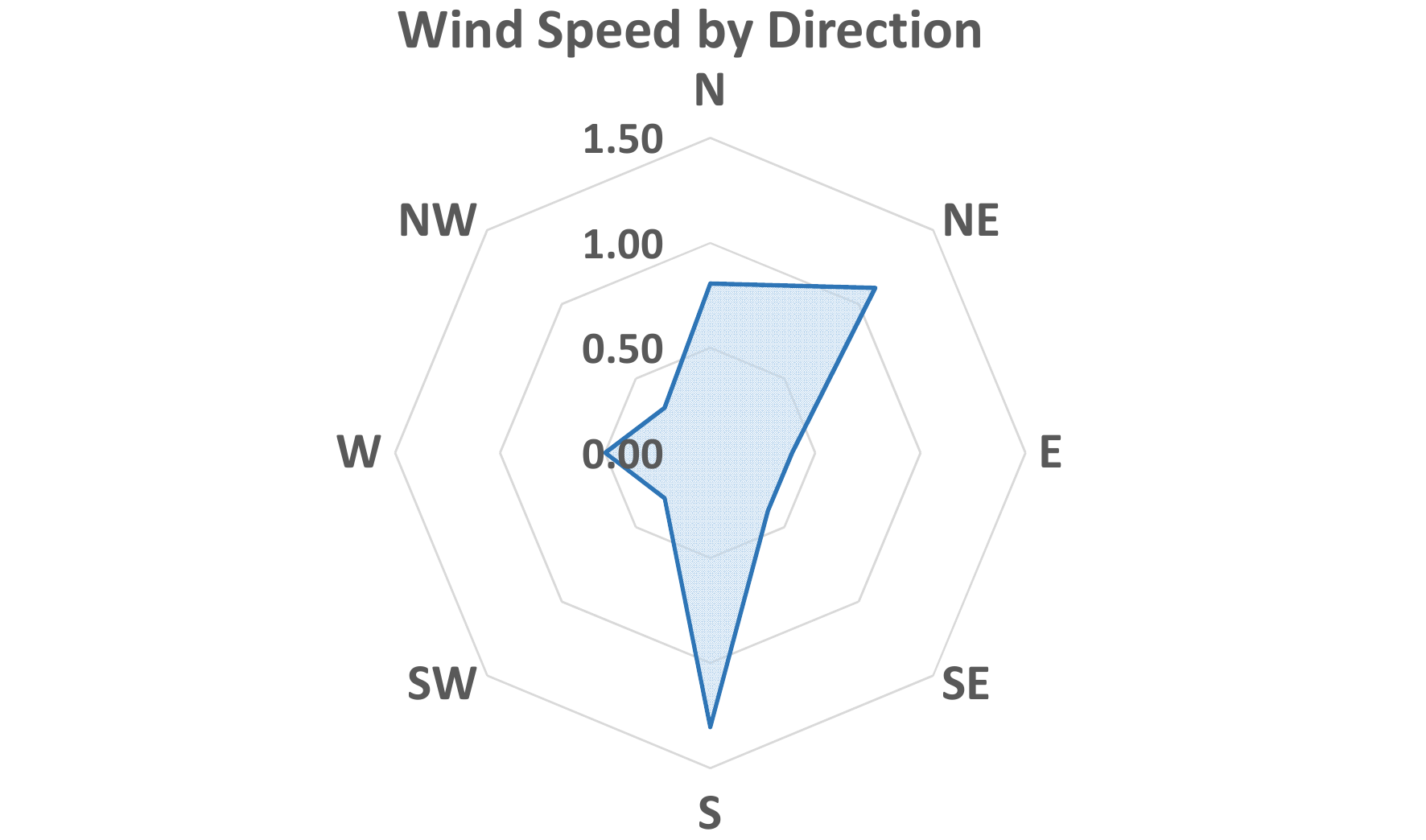}\label{fig:radar}}\\
	\caption{Example Charts for \reftab{tab:A} and \reftab{tab:B}. \label{fig:chart-examples}}
\end{figure*}

Creating charts for a multi-dimensional dataset (denoted as table) is a common activity in many domains such as education, research, engineering, finance, \etc To discover insights and perform routine analysis, people spend a huge amount of time constructing different types of charts to present diverse perspectives on their tables -- such as the charts in \reffig{fig:chart-examples} created for Table \ref{tab:A} and \ref{tab:B}.
Both \textbf{data queries} (selecting \textit{what} data to analyze) and \textbf{design choices} (\textit{how} to visualize selected data) are made during chart creation~\cite{hu2019vizml}. This tedious process requires experience and expertise in data analytics and visualization tools. For example, to compose the bar chart in \reffig{fig:bar-example}, one has to first select the left-most three fields/columns from \reftab{tab:A}, then choose bar chart type, map the three fields onto x and y axis, stack two value series one upon another, \etc

\begin{figure}[!h]
	\centering
	\includegraphics[width=\columnwidth]{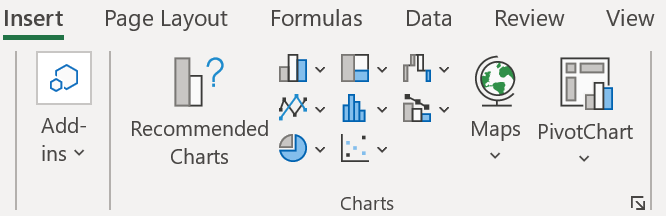}
	\caption{An Example of Chart Creation Entry UI.\label{fig:ExcelUI}}
\end{figure}

\begin{table*}[tbp]
	\centering
	\caption{Two Example Tables.\label{fig:ExampleTables}}
	\subfloat[Student Statistics Table.\label{tab:A}]{\includegraphics[width=0.6\linewidth]{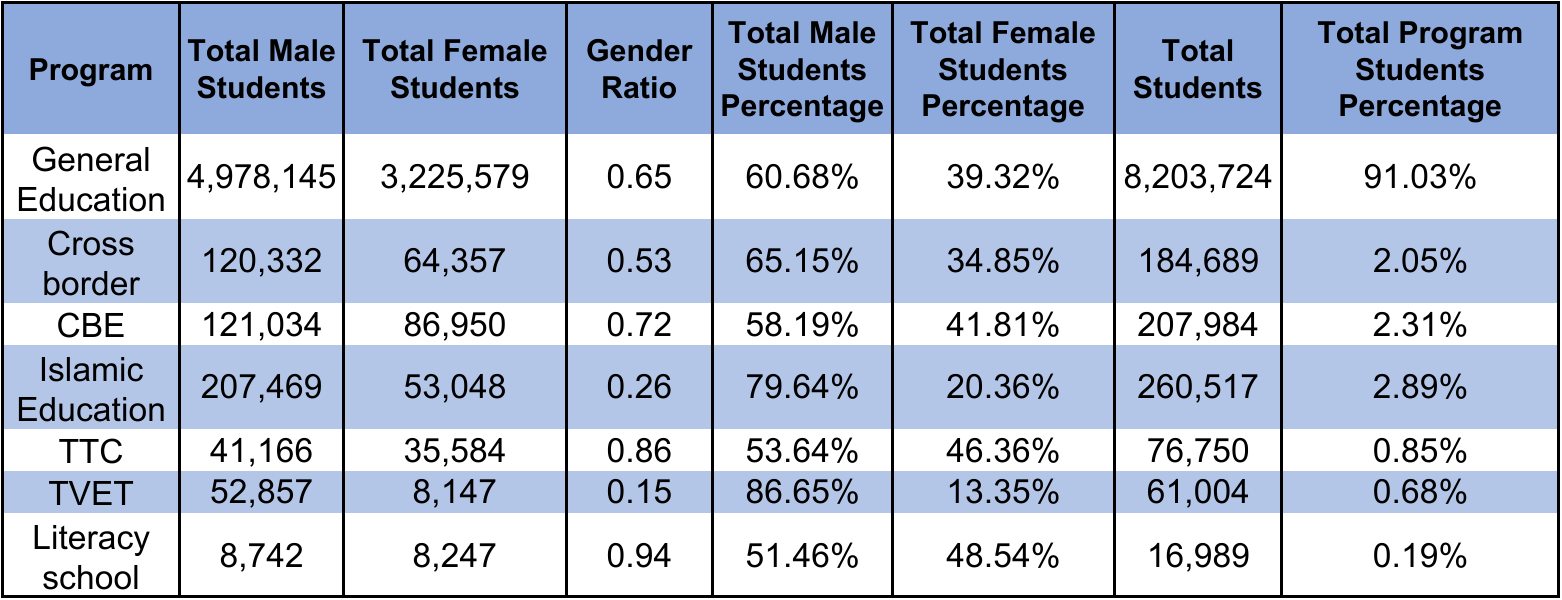}\label{fig:table-student}}\quad
	\subfloat[Evapotranspiration and Wind Table.\label{tab:B}]{\includegraphics[width=0.357\linewidth]{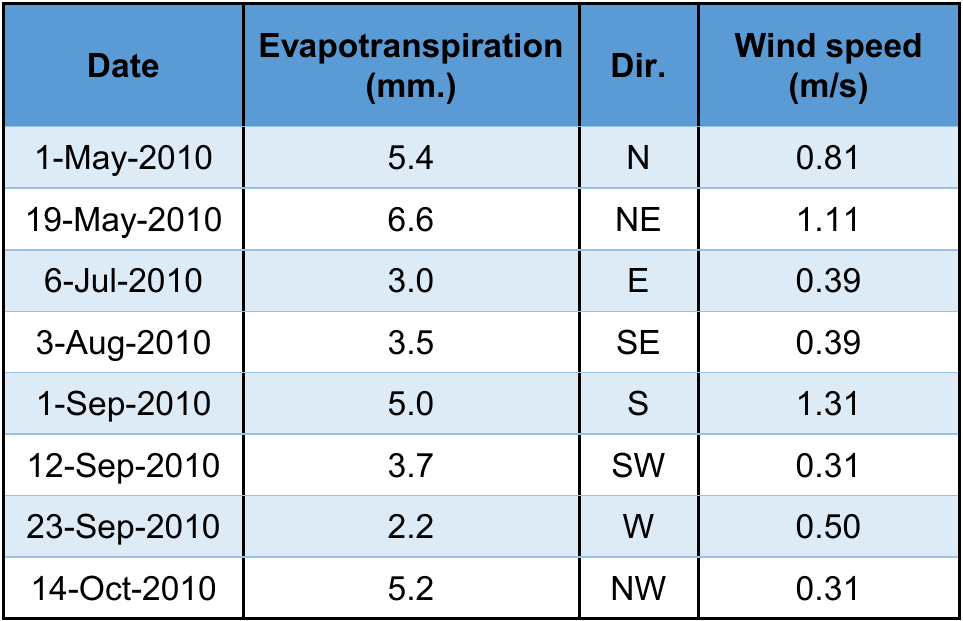}\label{fig:table-B}}\\	
\end{table*}

To simplify chart composing, a long line of works tried to build machine learning models recommending data queries and/or design choices, such as DeepEye~\cite{luo2018deepeye}, Data2Vis~\cite{dibia2019data2vis}, DracoLearn~\cite{moritz2019formalizing} and VizML~\cite{hu2019vizml}. However, most of them did not address the \textbf{single-type} tasks that each recommends one specific type of charts for a given table, including less used but meaningful minor chart types (\eg, area and radar charts). They only considered the \textbf{multi-type} task where a ranked list of few major types of charts (\eg, line, bar, scatter and pie charts) are recommended together.
But both single and multi-type tasks should be tackled for real-world scenarios: 
When facing a table for the first time, one usually has no clear idea about what chart should be created. In this scenario, an assistant could help leverage past common wisdom of what commonly composed charts could be created for the table -- which is the \textit{multi-type} task. For example, in Excel, the ``Recommended Charts'' button in \reffig{fig:ExcelUI} is expected for this.
Later with a clearer intention in mind, the main obstacle is the efforts needed to realize ideas through the complex charting process. Since lots of charting tools put chart type buttons / choices as the top entry points to chart composing (\eg, the chart type icons in \reffig{fig:ExcelUI}), guessing and suggesting auto-filling and completion of the details of a chosen chart type -- which is the \textit{single-type} task -- could help save time and efforts from users.
For example, when the first three fields of \reftab{tab:A} are selected, after clicking the bar chart icon on \reffig{fig:ExcelUI}, a lot of efforts could be saved if the rest of design choices on field mapping and stacking could be done automatically, leading to the bar chart in \reffig{fig:bar-example}.

When tackling the single-type and multi-type tasks with both data queries and design choices, there are three fundamental challenges. 
First, \textbf{separate costs}: It is memory and speed inefficient to design, train and deploy models for multi-type task and single-type tasks repeatedly and independently.
Second, \textbf{imbalanced data}: The available data for different chart types are highly imbalanced. Four major types of charts (line, bar, scatter and pie) cover $98.91\%$ of the available charts while others rarely appear because it is hard for non-experts to create them. Lack of data in minor chart types (area and radar) makes it hard to build high-quality models for them.
Third, \textbf{table as context}: Selecting and visualizing data from a table depend on not only the data statistics, but also the semantic meanings of the whole table context. Proper models need to be designed to take table context into chart recommendation.

In this paper, we propose \textbf{Table2Charts} framework to learn common patterns of chart creation -- including both data queries and design choices -- from a large amount of (table, charts) examples and to recommend charts for each given table. In \refsec{sec:problem}, charts recommendation is formulated as table to sequence(s) problem with next-action-token estimation to fill chart template(s). This formulation allows chart recommendation with partial intent when part of data queries and design choices are already given. 
Then in \refsec{sec:method}, as the estimation heuristic for beam searching, we design an encoder-decoder deep Q-value network (DQN) which selects table fields to fill template(s) via copying mechanism. All recommendation tasks share one encoder but have their own decoders, which addresses the separate costs challenge. The DQN is trained using mixed learning on the multi-type task of major chart types. By exposing its encoder part to the diverse source tables of different chart types, it learns \textbf{shared table representations} containing semantic and statistic information of table fields. Then the pre-trained table representations are transferred for type-specific decoders of single-type tasks, relieving the imbalanced data problem.

From the public web, we collect a large corpus of 266252 charts created from 165214 tables in Excel files and use a public Plotly corpus of 67617 charts from 36888 tables to verify the effectiveness of Table2Charts framework in \refsec{sec:exp}. 
For each chart type, the recall for top-3 and top-1 recommendations are $59.99\%\sim94.04\%$ and $49.30\%\sim79.72\%$ on the single-type tasks. The multi-type task of recommending major chart types has $61.84\%$ recall at top-3 and $43.84\%$ recall at top-1, which exceed the baseline methods whose maximal recall numbers are $27.14\%$ and $13.17\%$ respectively. Human evaluation is also conducted to validate the precision of the proposed framework on 500 frequently visited web tables from a search engine.
Lastly, through T-SNE visualization, we find that the DQN could learn shared table representations during multi-type task training for later transfer learning, thus improving the performance and saving memory occupation of single-type tasks.
All these experiments and evaluations justify that Table2Charts could efficiently learn to help composing charts.

In summary, our main contributions are:
\begin{itemize}
    \item Table2Charts framework is proposed by us to learn human chart composing wisdom. It generates both data queries and design choices in an action sequence for multi-type and single-type chart recommendation tasks with the state of the art performance and efficiency.
    \item To the best of our knowledge, we conduct the largest scale (165k tables and 266k charts from Excel corpus) training with diverse evaluations (on Excel, Plotly, and web table corpora) of chart recommend systems.
    \item We show the feasibility of learning shared table representations (encoding table fields into embedding vectors) for enhancing down-stream data analysis tasks.
\end{itemize}

\section{Problem}
\label{sec:problem}

To build machine learning models that learn patterns from large amounts of (table, charts) pairs, and to generate commonly composed charts for a given table, in this section we formulate single-type and multi-type chart recommendation tasks as table to action sequences generation by filling chart grammar templates.

A \textbf{table} here is an $n$-dimensional dataset $\mathcal{D}$ which contains $n$ data fields $\mathcal{F}_{\mathcal{D}} = (f_1^{\mathcal{D}},\cdots,f_n^{\mathcal{D}})$. Each data field refers to an attribute of the dataset with its corresponding header name (attribute metadata) and data values (records). For example, each column from tables in \reffig{fig:ExampleTables} is a data field with its first row as header.

To demonstrate our ideas, as shown in \reffig{fig:chart-examples}, in this paper we pick four \textbf{major} and two \textbf{minor chart types} that appeared in common charting tools such as Excel. The major types are line, bar, scatter and pie charts.
The minor types are area\footnote{Area Chart is very similar to line chart. The difference is that in area chart, the area between axis and line are commonly emphasized with colors or textures, so that the scale of color fill indicates the volumes. Commonly, area charts are used to represent accumulated totals using numbers or percentages over time.} and radar\footnote{Radar chart is used to compare the properties of a single component or the properties of two or more variables together.} charts.

\subsection{Chart Templates}
\label{sec:templates}

Although different types of charts exhibit distinct visual effects and behaviors, the essential actions for creating them from tables can be summarized into two categories: Selecting / referencing table fields and running specific charting commands / operations to organize and plot the selected fields. In this sense, a chart can be regarded as a sequence of actions on data queries and design choices.

\begin{definition}[Action Space / Tokens]
\label{def:tokens}
For an $n$-dimensional table $\mathcal{D}$, there are two categories of action tokens $\mathcal{A_D}=\mathcal{F_D}\cup\mathcal{C}$ representing core actions of composing a chart:
\begin{itemize}
    \item \textit{Field referencing token $f\in\mathcal{F_D}$} that indicates a field is selected for composing chart.
    \item \textit{Command tokens} (denoted as $\mathcal{C}$) which defines other commands for structuring a chart, including:
    \begin{enumerate}
        \item Chart type tokens, such as $\texttt{[Line]}$ means to start composing a line chart sequence;
        \item Separator \texttt{[SEP]} which splits the referenced fields with different roles in a chart sequence;
        \item Group operations in $\mathcal{G}=\{\texttt{[Cluster]}, \texttt{[Stack]}\}$\footnote{$\texttt{[Cluster]}$ means the values from several fields are put side-by-side, while $\texttt{[Stack]}$ means accumulating them one-upon-another for each x category / label (\Eg, \reffig{fig:bar-example}).} indicating how to put multiple data values from multiple fields (series) together along the x axis.
    \end{enumerate}
\end{itemize}
\end{definition}

Then we can define how to represent different types of charts using these unified action tokens. Unlike flexible language modelling in NLP, here action tokens should be organized into a sequence according to specific grammar rules of a chart type.

\begin{definition}[Chart Grammar Templates]
\label{def:templates}
The grammar templates of each chart type can be defined in the Backus-Naur form:
\begin{bnf*}
    \bnfprod{Line} {\bnfts{\texttt{[Line]}} \bnfpn{f+} \bnfts{\texttt{[SEP]}} \bnfpn{f*}  \bnfts{\texttt{[SEP]}}}\\
    \bnfprod{Bar} {\bnfts{\texttt{[Bar]}} \bnfpn{f+} \bnfts{\texttt{[SEP]}} \bnfpn{f*}  \bnfpn{grp}}\\
    \bnfprod{Scatter} {\bnfts{\texttt{[Scatter]}} \bnfpn{f} \bnfts{\texttt{[SEP]}} \bnfpn{f} \bnfts{\texttt{[SEP]}}}\\
    \bnfprod{Pie} {\bnfts{\texttt{[Pie]}} \bnfpn{f} \bnfts{\texttt{[SEP]}} \bnfpn{f*}  \bnfts{\texttt{[SEP]}}}\\
    \bnfprod{Area} {\bnfts{\texttt{[Area]}} \bnfpn{f+} \bnfts{\texttt{[SEP]}} \bnfpn{f*} \bnfts{\texttt{[SEP]}}}\\
    \bnfprod{Radar} {\bnfts{\texttt{[Radar]}} \bnfpn{f+} \bnfts{\texttt{[SEP]}} \bnfpn{f*} \bnfts{\texttt{[SEP]}}}
\end{bnf*}
where $\bnfpn{grp}$, $\bnfpn{f*}$, $\bnfpn{f+}$ and $\bnfpn{f}$ are token placeholders:
$\bnfpn{grp} \bnfpo$ an operation $\in \mathcal{G}$,
$\bnfpn{f*} \bnfpo \bnfes \bnfor \bnfpn{f} \bnfpn{f*}$,
$\bnfpn{f+} \bnfpo \bnfpn{f} \bnfor \bnfpn{f} \bnfpn{f+}$,
and $\bnfpn{f} \bnfpo \bnftd{a field} \in \mathcal{F_D}$, $\bnfes$ means empty.
The first $\bnfpn{f+}$ or $\bnfpn{f}$ segment is the \textbf{y-field}(s) and the second $\bnfpn{f*}$ or $\bnfpn{f}$ segment is the \textbf{x-field}(s).~\footnote{\textbf{X-fields} are the fields mapped to x-axis in line, bar, scatter and area charts, to legend in pie charts, and to curved polar axis in radar charts. Multiple x-fields means concatenation. \textbf{Y-fields} are the fields mapped to y-axis in line, bar, scatter and area charts, to the size of slice in pie charts, and to radial axis in radar charts. Multiple y-fields means multiple value series are shown together.}
Note that how x and y axes behave also depends on chart type: \Eg, scatter and pie charts only allow one y-field; temporal records will be ordered by their timestamps along x-axis on a line chart; \etc

Hard constraints are also included in the template definitions to restrict heuristic beam searching (see \refsec{sec:method}). These could be any hand-written rules, such as the data type of a field mapping to y-axis is forbidden to be string type. Currently we only set field type and field number limitations and let Table2Charts models to learn the rest. More rules such as the ones in Draco~\cite{moritz2019formalizing} could be adopted as hard constraints to further improve the framework.
\end{definition}

With the above definitions, now each chart can be written down as an action sequence. For example, the sequence of the bar chart in \reffig{fig:bar-example} (created from \reftab{tab:A}) is \texttt{[Bar]} (Total Male Students) (Total Female Students) \texttt{[SEP]} (Program) \texttt{[Stack]}.

There are more detailed charting aesthetics~\cite{wilke2019fundamentals} to consider, such as shape, size, color, line width and type, \etc In this paper, rather than considering every detail, we mainly focus on the core parts of data queries and design choices -- how to select and compose fields as axes of proper chart type -- to study if Table2Charts framework can learn common wisdom in a data-driven way.
Meanwhile, for the sake of simplicity, we only deal with database-like tables and referencing a whole field without filtering, aggregation, bucketing or ordering.
All the above subtle aspects of analysis may not be well supported by our training data (Excel files on the public web, see \refsec{sec:dataset}) in both quantity ($<5\%$ charts involve customizing them) and quality (the creators of these files may not be experts on inessential parts of charting). They can still easily be added as new tokens into the action space and grammar templates in the future.

\subsection{Table to Sequence Generation}
\label{sec:table2seq}
Table to charts recommendation now becomes how to meaningfully fill the placeholders of chart template(s). In other words, how to learn common wisdom to generate action token sequences (token-by-token from left to right) that follow the grammars of the given template(s).
Note that in a single-type task the first chart type token is fixed (the generation starts from the second token), while in the multi-type task it starts from the first chart type token.

A common way to solve sequence generation problem is to learn an estimation function for heuristic beam searching (more details in appendix \refsec{app:beam-search}).
Given a table $\mathcal{D}$ and an incomplete chart sequence $s$, we can define the valid actions space of the sequence as $\mathcal{A_D}(s)$ according to its corresponding template.
Follow the language modelling formulation in \cite{Zhou:2020:Table2Analysis}, we choose $Q(s,a)=P(sa \in \mathcal{T_D^+} \mid s, \mathcal{D})$ as action-value function to guide the choice of next action token $a\in \mathcal{A_D}(s)$. Here $\mathcal{T_D^+}$ is the set of all target chart sequences (the charts that would be adopted by user for $\mathcal{D}$) and their prefixes. So the optimal action-value function
$
    q_*(s, a)=\begin{cases}
    1 & \text{if }s'=sa\text{ and }s'\in \mathcal{T_D^+},\\
    0 & \text{otherwise}.
    \end{cases}
    \label{equ:Qtarget}
$
is the learning target for $Q(s,a)$. More details of the corresponding Markov decision process can be found in appendix \refsec{app:MDP}.


\section{Method}
\label{sec:method}

\begin{figure}
	\centering
	\includegraphics[width=\columnwidth]{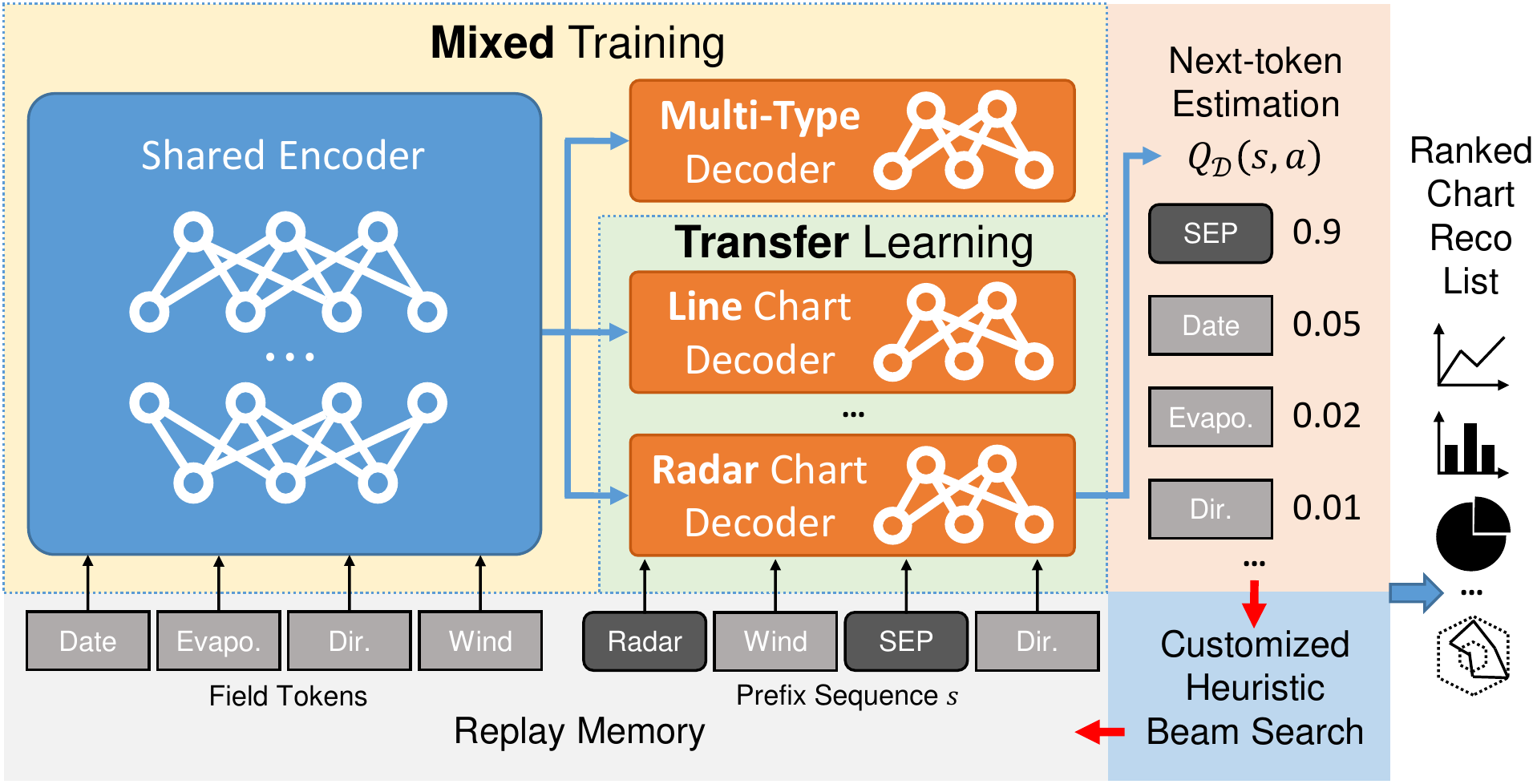}
	\caption{Overview of Table2Charts Framework.}
	\label{fig:overview}
\end{figure}

An overview of Table2Charts framework is shown in \reffig{fig:overview}.
To approximate $q_*(s, a)$ for chart generation, in \refsec{sec:DQN} we design an encoder-decoder deep Q-Network (DQN) architecture with copying mechanism.
Because the exposure bias is severe for sequence generation with templates, in \refsec{sec:training} we adopt search sampling technique to train DQN during beam searching.
Finally, in order to solve the imbalanced data problem and mutually enhance the performance among different chart types, in \refsec{sec:MTL} we propose a mix-and-transfer training paradigm for all the single and multi-type tasks.

\subsection{Filling Templates: DQN with Copying}
\label{sec:DQN}

\begin{figure*}
\subfloat[Customized CopyNet Overview.\label{fig:CopyNet}]{\includegraphics[width = 0.71\linewidth]{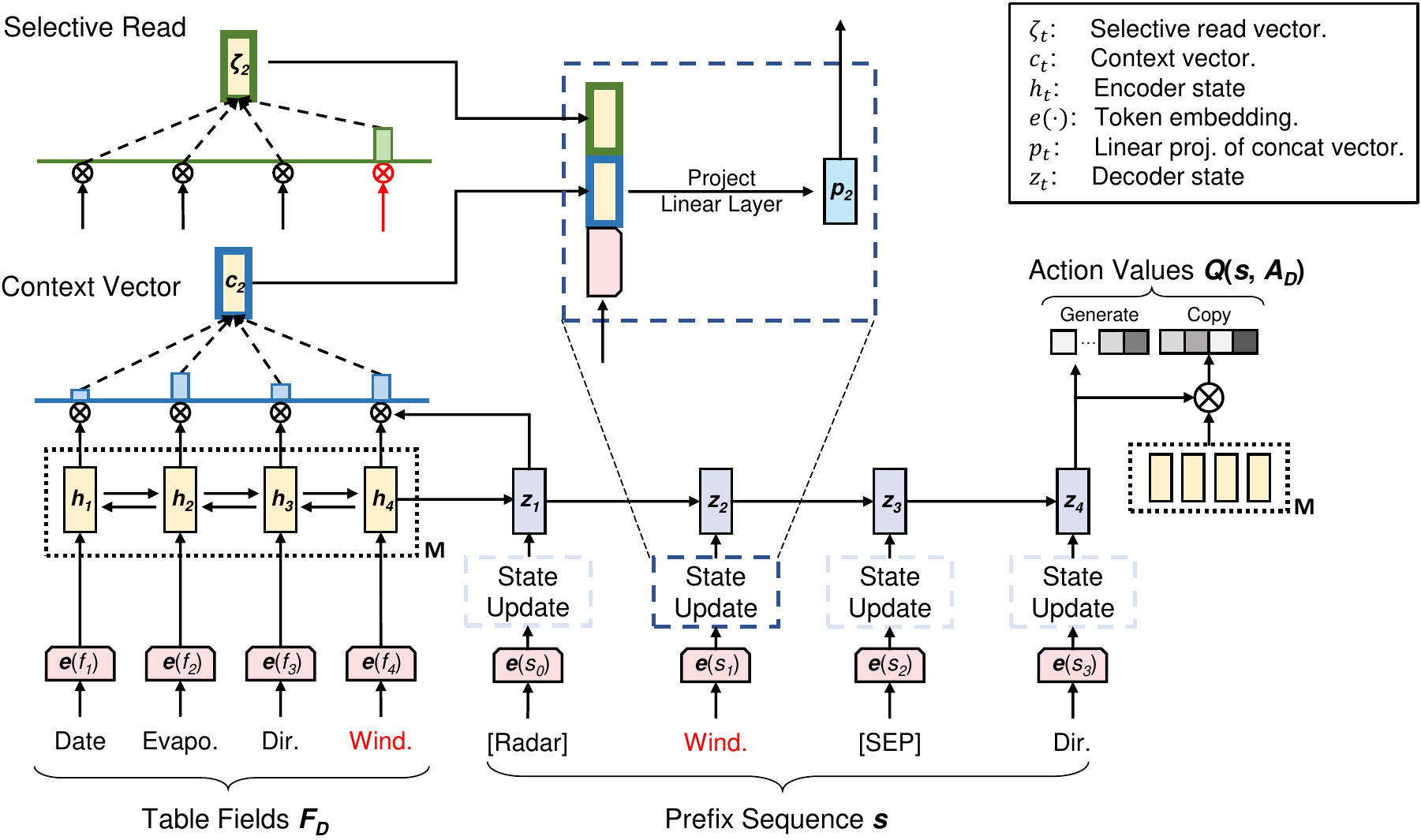}}
\quad
\subfloat[Token Embedding $e(\cdot)$.\label{fig:InputEmbed}]{\includegraphics[width = 0.25\linewidth]{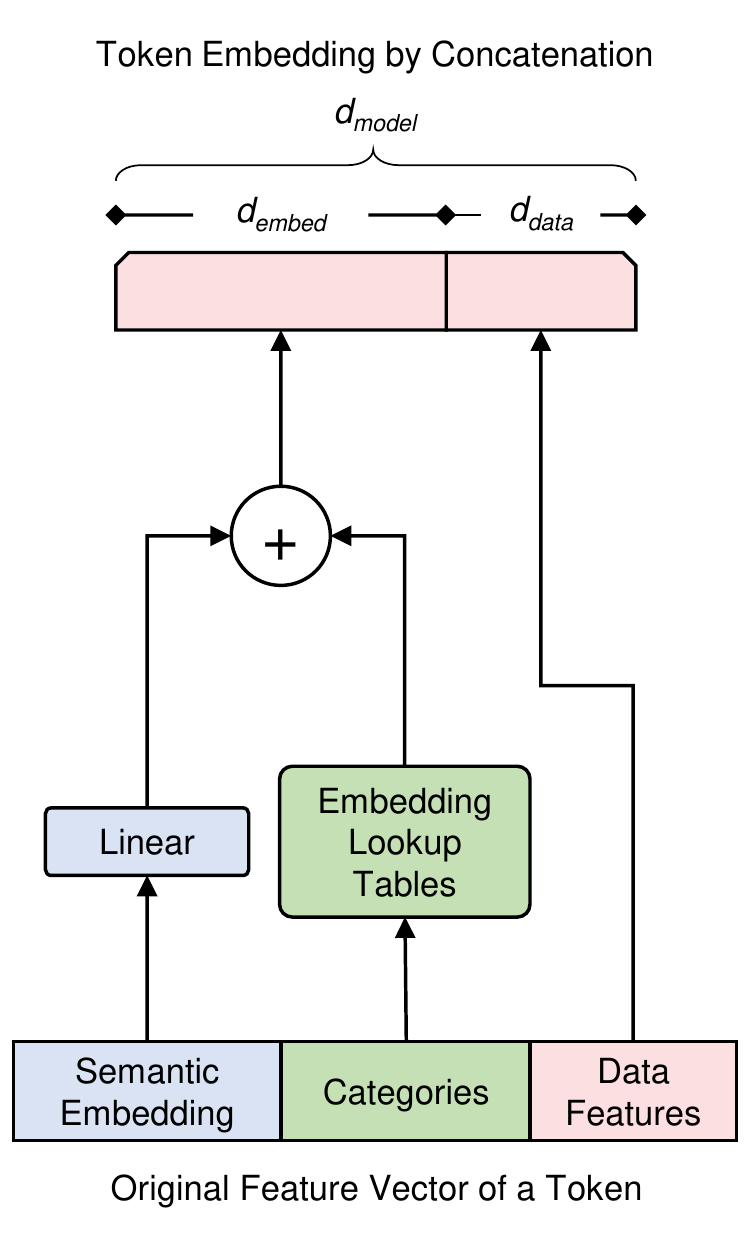}}
\caption{$q_*(s, a)$ Approximator: DQN Model Architecture.\label{fig:DQN}}
\end{figure*}

As shown in \reffig{fig:DQN}, we design a DQN (deep Q-network) $Q(s, \mathcal{A_D})$ to approximate $q_*(s, a)$. $Q(s, \mathcal{A_D})$ takes all the fields $\mathcal{F_D}=(f_1, ..., f_n)$ and a state $s=s_0...s_{T-1}$ as its input, and calculates the estimated action values ($\in [0,1]$) for all $a\in \mathcal{A_D}$. Only the outputs for $\mathcal{A_D}(s)$, \ie the valid actions \wrt the template grammar of $s$, are considered.

In \reffig{fig:CopyNet} is our customized CopyNet architecture. As shown in \reffig{fig:CopyNet}, the output vector of $Q(s, \mathcal{A_D})$ consists of two parts: ``Generate'' (for the command tokens) and ``Copy'' (for the field tokens). The ``Generate'' part contains the action value estimations for command tokens $\mathcal{C}$, which comes from a full connected layer with a binary softmax applied on the final decoder state $z_T$. The ``Copy'' part has variable length of value estimations for $\mathcal{F_D}$, which comes from a binary softmax applied on the product of $z_T$ and a non-linear transformation of the memory $M=\{h_1, ..., h_n\}$ (the encoder outputs).
We adopt GRU~\cite{cho2014learning} in bidirectional and unidirectional ways for the encoder and decoder RNN parts, respectively.
Thus $M$ is simply the outputs of a bidirectional GRU for $\mathcal{F_D}=(f_1, ..., f_n)$. A decoder state $z_t$ is updated by $p_t$ from the previous state $z_{t-1}$ in GRU cell. $p_t$ is a linear projection from the concatenated vector of three parts: selective read vector $\zeta_t$, context vector $c_t$ and the token embedding $e(s_{t-1})$. Selective read vector\footnote{Since each field token in $s$ can only refer to one unique field from $\mathcal{F_D}$ rather than possibly multiple source tokens in the original CopyNet, the calculation of selective read vector $\zeta$ is simplified in $Q(s, \mathcal{A_D})$.} choose the field representation from $M$ for a field token:
$
    \zeta_t = \begin{cases}
    h_\tau & f_\tau=s_{t-1},\\
    \vec{0} & \text{otherwise}.
    \end{cases}
$
Context vector $c_t$ is a linear attention between $z_{t-1}$ and $M$:
$
    c_t = \sum_{\tau=1}^{n}{\frac{e^{\eta(z_{t-1}, h_\tau)}}{\sum_{\tau'}{e^{\eta(z_{t-1}, h_{\tau'})}}} h_\tau}
$
where $\eta(\cdot, \cdot)$ is a linear function on two vectors.

The token embedding part omitted in \reffig{fig:CopyNet} is shown in \reffig{fig:InputEmbed}. It is part of encoder and is shared with decoder. Three kinds of token features (details in \refsec{app:features}) are fused together: 1) Semantic embedding of header name using FastText~\cite{bojanowski2017enriching}; 2) Categories including token type, field data type, \etc; 3) Data features about the statistics and distribution information of data values.

There are several differences between our model and the original CopyNet architecture~\cite{gu2016incorporating}: 
First, unlike the typical NLP scenario where vocabulary size is far greater than the length of the copying source, vocabulary (the command tokens) in Table2Charts is small while the universe of table fields is infinite, and there is no overlapping between generate mode (for vocabulary) and copy mode (for table fields). 
Second, the input tokens to $Q(s, \mathcal{A_D})$ first go through the feature transformation network $e(\cdot)$ in \reffig{fig:InputEmbed} rather than the usual index to embedding matrix in NLP. 
Third, the output of $Q(s, \mathcal{A_D})$ is a vector of $[0,1]$ values for each action, rather than a probability distribution over all actions in the original CopyNet. 

Our design of DQN with copying mechanism is naturally suited for tasks generating structures from table fields. It handles the open vocabulary of table field universe and provides a clear division between table representation (encoder) and template filling (decoder). The encoder part takes in the whole \textbf{table context} and generate field embedding vectors as table representations. The decoder part consumes these vectors for next-token estimation. 
As good $q_*(s, a)$ estimator, $Q(s, \mathcal{A_D})$ is then used by Table2Charts as a heuristic function in beam searching to generate multiple sequences.

\subsection{Fixing Exposure Bias: Search Sampling}
\label{sec:training}
A traditional way to train a next-token estimator is through teacher forcing~\cite{williams1989learning} by only sampling the prefix sequences of user-created charts, and comparing the estimated actions with actual user actions. In other words, in teacher forcing the only samples used to train $Q(s, \mathcal{A_D})$ network come from a corpus of (table, charts) pairs following the format of $q_*(s, a)$ (see \refsec{sec:table2seq}) with $s\in \mathcal{T_D^+}$.

As discussed in \cite{ranzato2015sequence,Zhou:2020:Table2Analysis}, with only teacher forcing, the outcome model could face exposure bias problem which is common in sequence generation. During teacher forcing, the model is only exposed to the ground truth states (target prefixes); While at inference time it has only access to its own predictions. 
As a result, during generation it can potentially deviate quite far from the actual sequence to be generated, leading to a biased estimation.

To mitigate exposure bias, we take the search sampling approach in \cite{Zhou:2020:Table2Analysis} to close the gap between training and inference. Inspired by reinforcement learning, the search sampling process adopts $Q(s, \mathcal{A_D})$ as the heuristic function to conduct beam searching on each table (details in appendix \refsec{app:beam-search}). Then the expanded states (including negative samples, $s$ not in $\mathcal{T_D^+}$) will be stored in a replay memory for periodical update of $Q(s, \mathcal{A_D})$ itself. This process is very effective after the warm-up of the network with teacher forcing. Without search sampling, the model would perform poorly with the customized beam searching process limited by chart templates.

\subsection{Mixed Training and Transfer Learning}
\label{sec:MTL}

As discussed in \refsec{sec:intro}, for single and multi-type tasks, there exists separate costs and imbalanced data challenges.
As shown in \reffig{fig:overview}, our basic idea to solve the challenges is that the table representations (the encoder part) can be shared by several (one multi-type and six single-type) tasks. This exposes the encoder to diverse and abundant table field samples, and reduce the memory occupation and inference time for deploying models of the tasks.

To train the shared table representation encoder and the task-specific decoders, as shown in \reffig{fig:overview}, we propose a \textit{mix-and-transfer} paradigm containing two stages: 
1) \textbf{Mixed Training}: Mixing samples from all major chart types together and train one DQN model. Its mixed encoder will be transferred to the next stage, while the whole mixed DQN will be used for the multi-type recommendation task.
2) \textbf{Transfer Learning}: Take the mixed encoder from the previous stage and freeze its parameters. Then, for each single-type task, a new decoder is trained with the fixed shared encoder using only the data of this chart type.

Comparing to \textbf{Separate Training} where a whole DQN is trained for each single-type task (using only the data of that chart type), the mix-and-transfer paradigm in Table2Charts has the following advantages: First, smaller memory occupation and faster inference speed, because now DQN models for all tasks share one same encoder, while separate training still inefficiently holds one for each task. This addresses the \textbf{separate cost} challenge.
Second, the encoder is exposed to far more samples than each individual chart type can provide. This not only leads to better learning and generalization of the table representation (see \refsec{sec:table-repr-study} on how \textbf{table context} is represented), but also addresses the \textbf{imbalanced data} challenge so that only decoder part (which is small comparing to the larger encoder part) needs tuning for minor chart types.

\section{Experiments}
\label{sec:exp}

In \refsec{sec:dataset}, we first introduce the (table, charts) corpora which are used for training and evaluating Table2Charts and other baseline models. Then in \refsec{sec:MTL-eval} and \refsec{sec:shared-eval}, the performance of mix-and-transfer paradigm (discussed in \refsec{sec:MTL}) is evaluated for single and multi-type tasks. Further empirical studies are also discussed in \refsec{sec:empirical} and \refsec{sec:table-repr-study}.

The experiments are run on Linux machines with 24 CPUs, 448 GB memory and 4 NVIDIA Tesla V100 16G-memory GPUs. Each training consists of 30 epochs of teacher forcing on 1 node followed by 5 epochs of search sampling (see \refsec{sec:training}) on 8 nodes. For fair comparisons, all evaluations are done on 1 node with the same configuration. By default, all evaluation metrics reported in this section are averaged over 5 runs for experiments with randomness.

\subsection{Chart Corpora}
\label{sec:dataset}
Two corpora -- Excel and Plotly -- are used for training and evaluation. The Excel corpus is created by us to train and evaluate models, but some baseline models do not provide training scripts. To make fair comparisons, in \refsec{sec:MTL-eval} we also evaluate every model on a public Plotly corpus~\cite{hu2019vizml} without training or fine-tuning on it.

\subsubsection{Excel Corpus}\label{sec:excel-corpus}
Our chart corpus contains 113390 (42.59\%) line, 67600 (25.39\%) bar, 64934 (24.39\%) scatter, 17436 (6.55\%) pie, 1990 (0.75\%) area and 902 (0.34\%) radar charts. They are extracted using OpenXML~\cite{openxml} from Excel spreadsheet files crawled from the public web. Following data preparation steps are also taken:

1) \textit{Cell Reference Cleansing}. X-fields, y-fields and series names\footnote{\textbf{Series names} refer to the name and meaning of each y-field, which are usually displayed in chart legend.} are stored as location references to spreadsheet cells (even in another file), which may lead to inaccessible or invalid tables. Charts with these kinds of cell references are removed from the corpus.

2) \textit{Source Table Restoration}. In spreadsheets, a chart object has no reference to its source table. (Only direct cell references are saved.) To restore the region and structure of its source table, we implement a table detection algorithm~\cite{dong2019semantic} according to its cell references. A chart will be dropped if its references are not covered by any detected table, the series names are not in the table header region, or the y-field references are not in the table value region.

3) \textit{Combo Chart Splitting}. In the corpus all combo charts are split into simple charts. Several simple charts (even in different types) can be drawn into one combined plot -- \eg, draw a line chart over a bar chart. (Note that simple charts can still have multiple x-fields and y-fields.) In this paper, we focus on simple charts and leave recommendation of combo charts as future work.

4) \textit{Table Deduplication}. To avoid the ``data leakage'' problem that duplicated tables are allocated into both training and testing sets, tables are grouped according to their schemas\footnote{Two tables are defined to have the same \textbf{schema} if they have the same number of fields, and each field's data type and header name are correspondingly equal.}. Then within each group, same (table, chart) pairs are merged.

5) \textit{Down Sampling}. After deduplication, the number of tables within each schema is very imbalanced -- 0.23\% schemas cover 20\% of these tables. To mitigate this problem, we randomly sample at most 10 unique tables for each unique (schema, chart) pair.

After preparing the data, 266252 charts are remained in 165214 unique tables with 98588 different schemas. The schemas (with their tables and charts) are randomly allocated for training, validation and testing in the ratio of 7:1:2. 

\subsubsection{Plotly Corpus}
We also adopt the public Plotly community feed corpus~\cite{hu2019vizml} and sample 36888 tables with 67617 charts (22644 line charts, 20053 scatter charts, 24204 bar charts and 716 pie charts) from it for testing in \refsec{sec:MTL-eval}.
To extract (table, charts) pairs, following the data processing procedure in VizML~\cite{hu2019vizml}, we download the full corpus (205GB) and adopt data cleansing code from VizML to remove charts with missing data. 
Also, similar procedures of \textit{combo chart splitting}, \textit{table deduplication} and \textit{down sampling} are applied to the remaining (table, charts) pairs as in the Excel corpus.

\subsection{Evaluations on Multi-Type Reco Task}
\label{sec:MTL-eval}

As mentioned in \refsec{sec:MTL}, for multi-type task, a mixed DQN is first trained using samples of Excel major chart types. Then, this mixed-trained DQN is used as heuristic function for beam searching to generate a ranked list of major-type charts for each table. We compare Table2Charts framework on recall and precision with four baselines: 
DeepEye~\cite{luo2018deepeye}, Data2Vis~\cite{dibia2019data2vis}, DracoLearn~\cite{moritz2019formalizing} and VizML~\cite{hu2019vizml}.

\subsubsection{Baselines}
\label{sec:baselines}
\textit{DeepEye} (\url{https://github.com/Thanksyy/DeepEye-APIs}) provides two public models (ML and rule-based) without training scripts. Thus, we adopt its models without training on our Excel corpus. Because its ML approach works better than its rule-based one on Excel test set, only its ML results are reported in this paper.
\textit{Data2Vis} (\url{https://github.com/victordibia/data2vis}) model was originally trained on 11 tables with 4.3k charts. 
For fair comparison, we re-train its model using our larger Excel training set (see \refsec{sec:excel-corpus}) which is also used by Table2Charts.
\textit{DracoLearn} (\url{https://github.com/uwdata/draco}) provides inference API without training scripts. Again, we evaluate it without training on our Excel chart corpus. It differs from the other methods in that it needs human defined rules as constraints and focuses on searching for specified chart components that least violates them. Thus, in Draco we adopt its default rules and specify it to generate chart type, x-fields and y-fields.
\textit{VizML} (\url{https://github.com/mitmedialab/vizml}) formulates the design choices into five classification problems and does not recommend data queries. In other words, VizML lacks ability to recommend charts without field selections. Thus, after re-training the classification models on the Excel training set, we only test VizML performance on design choices.
More details of experiment setup can be found in appendix \refsec{app:exp_setups}.

\subsubsection{Large-scale Evaluations on Recall}
\label{sec:recall-eval}
Recall of user charting actions in three stages -- data queries, design choices and overall chart recommendation -- are evaluated on Excel (testing set) and Plotly (whole dataset) corpora for Table2Charts and the four baselines. On data queries, we examine whether the recommended fields match user-selected ones. On design choices, we evaluate whether models can recommend correct chart type, field mapping and bar grouping operation given the user-selected fields. On overall chart recommendation, both data queries and design choices are compared with the ground truth.
Recall at top-$k$ ($k=1,3$; R@1, R@3) numbers are adopted as evaluation metrics. They show how a ranked list of chart recommendations matches the user-created charts. More details of recall calculation can be found in appendix \refsec{app:related-work}.

\begin{table}[t]
    \centering
    \caption{Evaluations of Table2Charts and Baseline Methods on Multi-Type Reco Task. (Averaged over 5 runs.)}
    \label{tab:baseline-compare}
    \resizebox{\columnwidth}{!}{
    \begin{tabular}{*{7}{c}}
        \toprule
        Dataset & Stage & Recall & DeepEye & Data2Vis & VizML & Table2Charts\\
        \midrule
        \multirow{6}{*}{Excel} & \multirow{2}{*}{Data Queries} & R@1 & 34.33\% & 31.16\% & - & \textbf{64.96\%} \\
        & & R@3 & 47.13\% & 42.18\% & - & \textbf{77.88\%} \\\cline{2-7}\vspace{-8pt}\\
        
         & \multirow{2}{*}{Design Choices} & R@1 & 17.83\% & 26.26\% & 21.38\% & \textbf{57.69\%} \\
         & & R@3 & 20.35\% & 48.88\% & - & \textbf{77.59\%} \\\cline{2-7}\vspace{-8pt}\\
         
         & \multirow{2}{*}{Overall} & R@1 & 10.18\% & 13.17\% & - & \textbf{43.84\%} \\
         & & R@3 & 15.85\% & 27.14\% & - & \textbf{61.84\%} \\
         \midrule
         
         \multirow{6}{*}{Plotly} & \multirow{2}{*}{Data Queries} & R@1 & 49.99\% & 63.78\% & - & \textbf{83.34\%} \\
        & & R@3 & 62.80\% & 71.82\% & - & \textbf{92.02\%} \\\cline{2-7}\vspace{-8pt}\\
        
         & \multirow{2}{*}{Design Choices} & R@1 & 37.69\% & 13.15\% & 30.21\% & \textbf{40.17\%} \\
         & & R@3 & 37.70\% & 32.85\% & - & \textbf{55.57\%} \\\cline{2-7}\vspace{-8pt}\\
         
         & \multirow{2}{*}{Overall} & R@1 & 25.05\% & 16.42\% & - & \textbf{33.37\%} \\
         & & R@3 & 35.98\% & 33.96\% & - & \textbf{48.03\%} \\
        \bottomrule
    \end{tabular}}
\end{table}

The recall numbers are shown in \reftab{tab:baseline-compare}. We can see that Table2Charts outperforms the baseline methods for all three stages on both Excel and Plotly corpora. The overall R@1 and R@3 have reached 43.84\% and 61.84\% on Excel (33.37\% and 48.03\% on Plotly), which exceeds those of baselines by large margins (at least doubled on Excel). 
The recall numbers of data queries stage are higher on Plotly than Excel -- This is because in Plotly corpus, each table only contain fields which are used in corresponding chart, and thus lower the difficulty of selecting fields. 
In addition to the results in \reftab{tab:baseline-compare}, DracoLearn has R@1 < 1\% on all stages -- It needs human-defined rules as constraints and focuses on searching for charts that least violates them, which leads to weak generalization.

\subsubsection{Human Evaluation on Precision}
To evaluate the quality of recommended charts (precision\footnote{Precision numbers cannot be calculated from Excel and Plotly corpora because they only have user-created charts but there can be good charts not created by users.}), we build a labelling website to collect and compare ratings for the top-1 recommendations from Table2Charts, DeepEye and Data2Vis\footnote{VizML and DracoLearn are not included because VizML cannot recommend complete chart with data queries and DracoLearn has weak generative power.}. 500 unique HTML tables crawled from the public web are selected based on query-frequency in a search engine. 10 experts working on web-table visualization manually label in the following way: 
For a given table, the website shows the table content for reading. When an expert confirms understanding the table, 3 charts recommended by the 3 models will be rendered with the same visualization library and shown in random order anonymously. Three 1 to 5 integer ratings
(higher score indicates better chart) are then labelled by the expert. 
Additionally, the expert is asked to mark if the table is actually unsuitable for chart recommendation. For every (table, 3 recommendation) pair, we collected results from 3 experts to avoid labelling bias.

\begin{table}[t] 
	\caption{Summary of Human Evaluation Ratings\label{tab:human-eval-distribution}}  
	\resizebox{0.82\columnwidth}{!}{
		\begin{tabular}{llllllllll} 
 		\toprule 
 	 	Rating & 5 & 4 & 3 & 2 & 1 & Avg & $\geq$4 & $\geq$3 & $\leq$2\\ 
  		\midrule 
 		Table2Charts & 517 & 158 & 115 & 102 & 98 & 3.90 & 675 & 790 & 200  \\ 
 		Data2Vis & 309 & 178 & 167 & 125 & 211 & 3.25 & 487 & 654 & 336  \\ 
 		DeepEye & 312 & 166 & 139 & 137 & 236 & 3.18 & 478 & 617 & 373   \\ 
  \bottomrule 
 \end{tabular}} 
\end{table}

We filtered out the tables which marked as ``unsuitable for chart recommendation''\footnote{Determining whether a table is suitable for generating a chart is out of the scope of this paper and would be part of future work.}, and got the distribution of the ratings based on 330 tables left. As shown in \reftab{tab:human-eval-distribution}, Table2Charts has the highest average score, the largest amount of good charts (rating=5, rating$\geq$4, rating$\geq$3), and the smallest amount of bad charts (rating $\leq$2).
 
To check statistical significance, we further conduct Wilcoxon signed-rank test~\cite{wilcoxon1970critical} which is a non-parametric statistical hypothesis test used to compare two related or matched samples to assess whether their population mean ranks differ (\ie it is a paired difference test). 
At 95\% confidence level, when comparing Table2Charts with DeepEye and comparing Table2Charts with Data2Vis, both $p$-values from Wilcoxon test are less than 0.0001.
These results show that the recommended charts from Table2Charts have better quality than those from DeepEye and Data2Vis.

\subsubsection{Efficiency Comparison}
On average, Table2Charts only takes 12.14ms to generate chart recommendations for a table, while it costs DeepEye and Data2Vis 48.19ms and 210ms, respectively. In summary, Table2Charts outperforms baseline methods on both performance and efficiency.

\subsection{Evaluations on Single-Type Reco Tasks}
\label{sec:shared-eval}

After mixed training, as discussed in \refsec{sec:MTL}, the shared table representation encoder is taken and frozen for the training of six decoders for six single-type tasks. 
For comparison, the separate training (see \refsec{sec:MTL}) will generate an independent DQN model for each chart type with the same settings as the transfer learning. Also, the mixed DQN from \refsec{sec:MTL-eval} is directly tested on single-type tasks of major types.

\begin{table}[t]
    \centering
    \caption{Evaluations of Three Training Methods (\refsec{sec:MTL}) on Six Single-Type Tasks. (Averaged over 5 runs.)}
    \label{tab:mtl-eval}
    \resizebox{0.78\columnwidth}{!}{
    \begin{tabular}{*{6}{c}}
        \toprule
        \multicolumn{2}{c}{Type(s)} & Recall & Separate & Mixed & Transfer\\
        \midrule
        \multirow{8}{*}{Major Type} & \multirow{2}{*}{Line} & R@1 & 52.02\% & 52.53\% & \textbf{53.78\%}\\
        & & R@3 & 68.28\% & 68.58\% & \textbf{69.37\%}\\
        \cline{2-6}\vspace{-8pt}\\
        & \multirow{2}{*}{Bar} & R@1 & 56.56\% & 58.69\% & \textbf{60.25\%}\\
        & & R@3 & 70.34\% & 72.07\% & \textbf{73.14\%}\\
        \cline{2-6}\vspace{-8pt}\\
        &\multirow{2}{*}{Scatter} & R@1 & 51.73\% & 54.69\% & \textbf{56.48\%}\\
        && R@3 & 69.33\% & 68.96\% & \textbf{74.24\%}\\
        \cline{2-6}\vspace{-8pt}\\
        &\multirow{2}{*}{Pie} & R@1 & 73.60\% & 77.99\% & \textbf{79.72\%}\\
        && R@3 & 90.60\% & 93.12\% & \textbf{94.04\%}\\
        \midrule
        \multirow{4}{*}{Minor Type}&\multirow{2}{*}{Area} & R@1 & 27.48\% & -- & \textbf{49.30\%}\\
        && R@3 & 40.32\% & -- & \textbf{59.99\%}\\
        \cline{2-6}\vspace{-8pt}\\
        &\multirow{2}{*}{Radar} & R@1 & 49.90\% & -- & \textbf{71.77\%}\\
        && R@3 & 60.93\% & -- & \textbf{77.00\%}\\
        \bottomrule
    \end{tabular}}
\end{table}

The evaluation results are shown in \reftab{tab:mtl-eval}.
The mix-and-transfer paradigm (``Transfer'') has higher recall numbers than separate training (``Separate'') and mixed-only (``Mixed'') DQN for all chart types. Table2Charts could handle single-type tasks well by learning shared \textbf{table context} representations.
Considering the model size where the encoder and decoder parts are designed with 1.3M and 0.5M parameters (see appendix \refsec{sec:t2c-details}), ``Separate'' have 10.8M parameters while ``Transfer'' only has 4.3M parameters. In this way, Table2Charts reduces \textbf{separate costs} and improves efficiency for model deployment and inference.

As for minor chart types, there are huge gaps between the performance of ``Transfer'' and ``Separate'' in \reftab{tab:mtl-eval}.
With mix-and-transfer (``Transfer'') paradigm, on average R@1 and R@3 increase 21.84\% and 17.87\%. The main reasons are that the shared encoder could capture and extract the information of table context and field semantics, and the quantity of minor type charts are sufficient to train decoder which is of smaller size. Thus, as discussed in \refsec{sec:MTL}, the \textbf{imbalanced data} problem is overcome.

\subsection{Recommendation Case Studies}
\label{sec:empirical}
As an example, in this section we take \reftab{tab:A} (with user-created \reffig{fig:bar} and \ref{fig:area}) to conduct empirical studies.
When a user does not know where to start, the multi-type mixed model (\refsec{sec:MTL-eval}) recommends common types of charts. Its top-3 recommendations are:\\
{\small
1) [\texttt{Bar}](Total Male Students)(Total Female Students)[\texttt{SEP}](Program)[\texttt{Stack}] \\
2) [\texttt{Bar}](Total Male Students)(Total Female Students)[\texttt{SEP}](Program)[\texttt{Cluster}]\\
3) [\texttt{Bar}](Total Program Students Percentage)[\texttt{SEP}](Program)[\texttt{Cluster}]}

From the above results, we can see that the mixed model successfully recommends the bar chart in \reffig{fig:bar} as the top-1 result. Our model can identify $f_2^{\text{\ref{tab:A}}}$-$f_3^{\text{\ref{tab:A}}}$ as a group and use them to create a bar chart with two y-fields. Result 2) is the clustered form of the bar chart in result 1). Result 3) is also useful. Our model can identify $f_8^{\text{\ref{tab:A}}}$ should not be a measure in the $f_2^{\text{\ref{tab:A}}}$-$f_3^{\text{\ref{tab:A}}}$ group. The multi-type model tend to recommend what are commonly composed, thus may lack diversity (\eg, all the above results are bar charts). So one can also put single-type recommendations into the list. (We leave as a future work how to mix the results from multi-type and single-type models together for a balanced recommendation.)

When a user has chosen a specific chart type and needs auto-completion help, it is time to use single-type models (\refsec{sec:shared-eval}) for recommendations. In such scenario, our single-type models can recommend all as top-1 the area chart in \reffig{fig:area} and the bar chart in \reffig{fig:bar}.
Furthermore, the single-typed model can also recommend the pie chart in \reffig{fig:pie}, which is also meaningful to show the percentages but not originally created by the user.

\subsection{Exploring Table Representations}
\label{sec:table-repr-study}
\begin{figure}[tbp]
	\centering
	\includegraphics[width=\columnwidth]{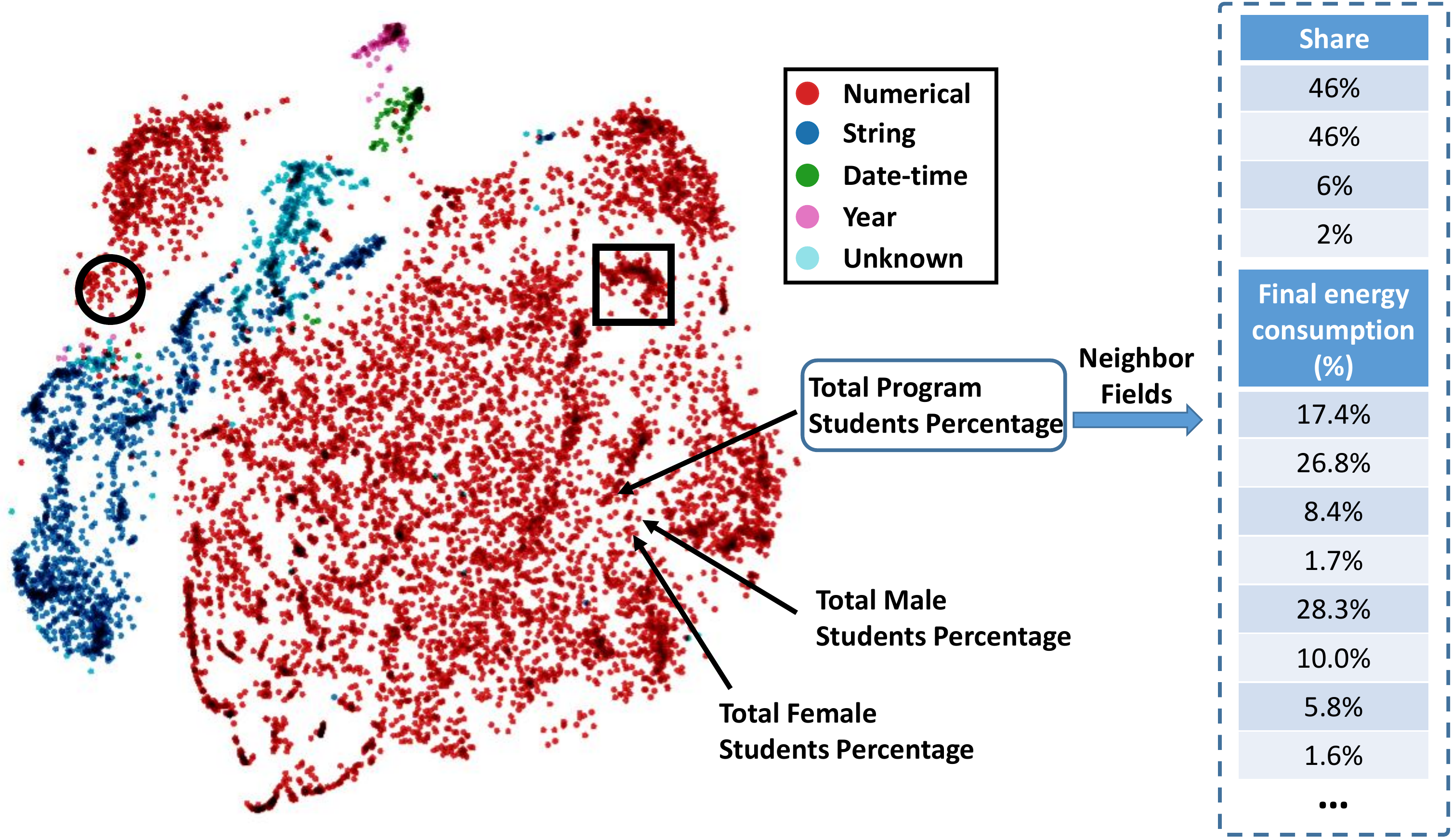}\\	
	\caption{Visualization of Shared Table Representations.}
	\label{fig:ExampleVisual}
\end{figure}

To understand how the embeddings generated by the shared table representation encoder work, from the validation set 20000 fields (from 3039 tables) are randomly chosen and visualized through t-SNE~\cite{smilkov2016embedding}.
In the left part of \reffig{fig:ExampleVisual}, each point represents a field and the color represents its field type.
In the figure, we can see the field type information is learnt by the embedding in a meaningful way. 
For example, date-time fields and year fields are close. One possible explanation is that they both are often used as x-axes in line charts, and thus have similar representations.

As depicted in \reffig{fig:ExampleVisual}, marked by arrows are the points corresponding to the fields $f_5^{\text{\ref{tab:A}}}$, $f_6^{\text{\ref{tab:A}}}$ and $f_8^{\text{\ref{tab:A}}}$ (which is ``Total Program Students Percentage'') from \reftab{tab:A}.
They are close to each other because their record values are all percentages. Note that $f_5^{\text{\ref{tab:A}}}$ and $f_6^{\text{\ref{tab:A}}}$ are closer compared to $f_8^{\text{\ref{tab:A}}}$ because their semantics are similar (contain gender information). Some example neighbor fields (based on cosine distance) of $f_8^{\text{\ref{tab:A}}}$ are shown in the right part of \reffig{fig:ExampleVisual}. Similar to $f_8^{\text{\ref{tab:A}}}$, these fields are also percentages that sum up to 1.

Two more clusters are shown as examples in \reffig{fig:ExampleVisual}.
In the squared area, there are many fields about countries. \Eg, there are four numerical fields (from four tables) with header names ``U.S.'', ``Japan'', ``England'' and ``Scotland'' showing annual statistics. 
In the circular area, many fields take the role of index or ID. 
\Eg, located in this cluster are four fields (from four tables) with header names ``VerminID'', ``Index'', ``category'' and ``Course Code'' and increasing integers. These integers lose the measure property -- they are not for mathematical operations / aggregations. Thus, these fields locate very close to string fields (dark blue points) in \reffig{fig:ExampleVisual}.

\section{Related Work}
\begin{table*}[t]
    \centering
    \caption{Comparisons among Different Chart Recommending Systems}
    \label{tab:system-compare}
    \resizebox{\textwidth}{!}{
    \begin{tabular}{*{8}{c}}
    \toprule
    System & Reco Tasks & Learning Approach & Models & Dataset & $N_{data}$ & Data Source & Data Generation\\
    \midrule
    Table2Charts& \makecell{Data queries +\\ Design choices} & \makecell{End-to-end chart generation \\as action token sequence} & \makecell{CopyNet as deep\\ Q-network} & \makecell{(full table, charts)\\ pairs} & \makecell{165k tables with\\ 266k charts} & Web Excel files & Human\\
    \midrule
    VizML & Design choices & 5 classification tasks & \makecell{Fully-connected \\ feed-forward NN} & \makecell{(partial table, charts)\\ pairs} & 119k tables & \makecell{Plotly community\\ feed} & Human \\
    \midrule
    DracoLearn & \makecell{Data queries +\\ Design choices} & \makecell{Soft constraints/rules weights\\ for clingo ASP solver} & RankSVM & Pairwise comparison & 1100 + 10 pairs & Various & \makecell{Rules $\rightarrow$\\ Annotations}\\
    \midrule
    Data2Vis & \makecell{Data queries +\\ Design choices} & \makecell{End-to-end chart generation \\as JSON string sequence} & \makecell{Character-level\\ seq2seq NN} & \makecell{(full table, charts)\\ pairs} & \makecell{11 tables with\\ 4300 charts} & \makecell{Tool\\ (Voyager)} & \makecell{Rules $\rightarrow$\\ Validations} \\
    \midrule
    DeepEye & \makecell{Data queries +\\ Design choices} & \makecell[l]{1. Good/bad classification\\2. Ranking} & \makecell[l]{1. Decision tree\\2. LambdaMART} & \makecell[l]{1. Good/bad chart labels\\2. Pairwise comparison} & \makecell{42 tables with\\1. 33.4k labels\\2. 285k pairs}  & Various &  \makecell{Rules $\rightarrow$\\ Annotations} \\ 
    \bottomrule
    \end{tabular}}
\end{table*}

\textbf{Analysis Recommendation}:
For general data analysis and insight recommendation from tables, current systems are mostly based on collaborative filtering~\cite{Marcel:2011:Survey}, statistical significance~\cite{Tang:2017:insights}, heuristic and history matching~\cite{ehsan2016muve,Milo:2018:NextStep}, or only target for specific analysis~\cite{Zhou:2020:Table2Analysis}. They rarely consider the semantic meaning of table context or tackle the challenges in recommending multiple types of analysis, which are both taken into account by Table2Charts in an end-to-end approach using large-scale human created corpus.

\textbf{Chart Recommendation} is an important branch in analysis recommendation. Lots of visualization recommendation systems heavily rely on hand-crafted heuristics and rules, such as Voyager~\cite{kanit2016voyager} and DracoLearn~\cite{moritz2019formalizing}. Data-driven approaches are becoming popular in recent learning-based systems such as DeepEye~\cite{luo2018deepeye}, Data2Vis~\cite{dibia2019data2vis} and VizML~\cite{hu2019vizml}.
In \refsec{sec:MTL-eval}, we discussed DeepEye, Data2Vis, DracoLearn and VizML as baselines. More of their differences with Table2Charts are summarized in \reftab{tab:system-compare}.
DracoLearn and DeepEye both learnt from low quality data and depended on complex rule designs. Data2Vis suffered from naive model of character-level seq2seq generation of Vega-lite~\cite{2017-vega-lite} JSON string. VizML only considered design choices without handling data queries.

\textbf{Structured Prediction}:
Filling chart templates and generating action sequences is a structured prediction problem. There are lots of related work such as NL QA and Text2SQL~\cite{wang-etal-2020-rat}. Table2Charts is inspired by \cite{gu2016incorporating,Zhou:2020:Table2Analysis} to design an encoder-decoder architecture with copying mechanism as a function approximator.

\textbf{Representation Learning and Pre-training}:
The word embedding~\cite{bojanowski2017enriching} and pre-training paradigm~\cite{Devlin:2018:BERT} in NLP inspired us to learn pre-trained table representations for multiple tasks~\cite{Zhang:2017:MTLSurvey}. The table field embedding could be useful for more down-stream data analysis tasks including recommendation of other types of analysis.

\section{Conclusion}
In this paper, we propose the Table2Charts framework to solve single and multi-type chart recommendation tasks considering both data queries and design choices. Through copying from table fields, shared table representations are learnt to enhance performance and efficiency for all chart types. We believe the proposed techniques can be widely used for data analysis tasks on tables in the future.

\balance
\bibliographystyle{ACM-Reference-Format}
\bibliography{references}

\clearpage
\begin{appendices}
\section{Table2Charts Framework Details}
\label{app:model-sup}
In this section, we will dive into the details of the table-to-sequence problem formulation in \refsec{sec:table2seq} with the corresponding Markov decision process. 
Also, the input token features to the input embedding network of \reffig{fig:InputEmbed} in \refsec{sec:DQN} will be listed for your references. Finally, we will describe the companion heuristic beam searching algorithm to the DQN. 
Core code of Table2Charts and part of the test data can be found at \url{https://github.com/microsoft/Table2Charts}.

\subsection{Markov Decision Process (MDP)}
\label{app:MDP}

As described in \refsec{sec:table2seq}, the MDP for chart generation is based on the one for pivot table that was first defined by \cite{Zhou:2020:Table2Analysis}.

\begin{definition}[Chart Generation MDP]
\label{def:MDP}
For a table $\mathcal{D}$, we adopt the definitions in \refsec{sec:templates} and \refsec{sec:table2seq} to describe the next-token chart sequence generation MDP:
\begin{itemize}[fullwidth,itemindent=2em]
    \item State space is $\mathcal{S_D^+} = \{s \mid s \in \bigcup_{l=1}^{\infty}{\mathcal{A_D}^l}, s\text{ is legal}\}$, which can be viewed as a forest with chart type tokens (see \refsec{sec:templates}) as root nodes (initial states) of the trees. $\mathcal{S_D^+}$ contains all the prefixes of all the possible legal chart sequences that follow the chart templates.
    \item Action space $\mathcal{A_D}$: 
    The legal actions for a given state $s$ are $\mathcal{A_D}(s) = \{a \mid sa\in \mathcal{S_D^+}, \forall{a\in \mathcal{A_D}}\}$.
    \item State transition is deterministic. The transition probability from $s$ to $s'$ by taking action $a\in \mathcal{A_D}(s)$ is: 
    $$
        P_\mathcal{D}(s,a,s') = \begin{cases}
        1 & \text{if }s'=sa,\\ 
        0 & \text{otherwise}.
        \end{cases}
    $$
    \item Reward function $R_\mathcal{D}$ is designed to reflect if a user-created sequence is successfully generated:
    $$
        R_\mathcal{D}(s,a,s') = \begin{cases}
        1 & \text{if }s'=sa\text{ and }s'\in \mathcal{G_D},\\
        0 & \text{otherwise}.
        \end{cases}
    $$
    Here $\mathcal{G_D}$ is a subset of $\mathcal{S_D^+}$ which contains exactly the charts created by user for $\mathcal{D}$. (As first mentioned in \refsec{sec:table2seq}, $\mathcal{T_D^+}$ is the set of all the prefixes of all the target sequences in $\mathcal{G_D}$.)
    \item Discount rate $\gamma = 1$ so that the length of a chart sequence has no impact on its rewards.
\end{itemize}
\end{definition}

According to Bellman optimality equation, one can easily find the optimal action-value function (the expected discounted return for the optimal policy):
$
    q_*(s, a) =R_\mathcal{D}(s,a,sa)+\gamma \max_{a'\in \mathcal{A_D}(sa)}{q_*(sa, a')}\\
    = \begin{cases}
    1 & \text{if }s'=sa\text{ and }s'\in \mathcal{T_D^+},\\
    0 & \text{otherwise}.
    \end{cases}
$
In other words, $q_*(s, a)$ equals to $1$ if and only if $sa$ is a prefix of a target sequence. As described in \refsec{sec:table2seq}, the rest of the problem is to learn a good approximator for
$q_*(s, a)$.

\subsection{Token Features for Input Embedding}
\label{app:features}
As shown in \reffig{fig:InputEmbed}, token embedding vector consists of:

\textbf{Semantic Embedding.} Semantic embedding features are calculated from the header name of a field (\eg, table header or database attribute string). In this work, we adopt FastText~\cite{bojanowski2017enriching} with $vocab size=200,000$ and $embed size=50$ for semantic embedding. If there are more than 1 words in the field name, the embedding of all words are averaged. 

\textbf{Field Categories.} There are five types of categorical features which are adopted in this work.
\begin{enumerate}
    \item \textit{Token type} shows the type of a token in an analysis sequence, which includes \{\texttt{PADDING}, \texttt{SEP}, \texttt{FIELD}, \texttt{GRP}, \texttt{Line}, \texttt{Bar}, \texttt{Scatter}, \texttt{Pie}, \texttt{Area}, \texttt{Radar}\}.
    \item \textit{Segment type} shows to which segment a token belongs in an analysis sequence. This categorical feature can be \{\texttt{PADDING}, \texttt{X}, \texttt{Y}, \texttt{GRP}, \texttt{OP}\}. \texttt{OP} corresponds to \texttt{SEP} and chart type tokens.
    \item \textit{Field type} shows the type of a field, which includes \{\texttt{Unknown}, \texttt{String}, \texttt{Year}, \texttt{DateTime}, \texttt{Decimal}\}.
    \item \textit{Field role} shows whether a field could be one of the left headers of a cross table (detected during source table restoration). The options include \{\texttt{Invalid}, \texttt{Header}, \texttt{Value}\}.
    \item \textit{Grouping operation} corresponds to $\bnfpn{grp}$, which includes \{\texttt{Invalid}, \texttt{Cluster}, \texttt{Stack}\}.
\end{enumerate}

\textbf{Data features.} 
We adopt the 16 statistic features in \cite{Zhou:2020:Table2Analysis}, and design 15 new features: SumIsIn01, SumIsIn0100, Range, Variance, Covariance, AbsoluteCardinality, MedianLength, LengthStdDev, AvgLogLength, ArithmeticProgressionConfidence, GeometricProgressionConfidence, Skewness, Kurtosis, GiniCoefficient, NRows.

All features except AvgLogLength are calculated for numerical fields. All applicable features are calculated for string fields. Most data features are ranged in $[0, 1]$, and for those whose range may be very large, we normalized them by their \(99\)th percentile numbers in the Excel chart corpus. Data statistic features for non-field tokens remain empty (their values are assigned as zeros).

\subsection{Heuristic Beam Searching}
\label{app:beam-search}
In search sampling training and beam searching inference stages, we adopt and customize a drill-down beam searching algorithm~\cite{Zhou:2020:Table2Analysis}. It takes the following steps to generate chart sequences:
\begin{enumerate}
    \item Initially, the searching frontier only contains the sequence(s) that each consists of one specified chart type token from $\{\texttt{[Line]},$ $\texttt{[Bar]}, \texttt{[Scatter]}, \texttt{[Pie]}, \texttt{[Area]}, \texttt{[Radar]}\}$. Chart types are chosen according to the training or inference task\footnote{In separate training, transfer learning and single-type inference, only one chart type is used; while in mixed training and multi-type inference, all major types are used.}.
    \item For each round, the top-$BeamSize$ scored partial / incomplete sequences in the frontier will be popped and extended as described below.
    \begin{enumerate}
        \item For each state in the beam, greedily drill down (choose $a$ with the highest $Q(s, a)$ to append) until a complete sequence is generated. The complete sequence is put into the result ranking list with $Q(s, a)$ as its score. Each non-optimal state $sa\ (a \in \mathcal{A_D}(s))$ from each expansion (one calculation of $Q(s, \mathcal{A_D})$) is put into the frontier also with $Q(s, a)$ as its score. \label{drill-down}
        \item No more rounds and stop searching if the number of expansions exceeds $ExpandLimit$.
    \end{enumerate}
\end{enumerate}

As mentioned in \refsec{sec:templates}, to restrict heuristic beam searching and eliminate some nonsense recommendations, hard constraints are defined in chart templates. For example, the data type of a y-field is forbidden to be string type. During training and inference, these hard constraints are also applied by removing illegal actions from $\mathcal{A_D}(s)$ for each expansion in the heuristic beam searching.

This also allows users to specify more constraints on searching. For example, a user could select interested fields from a table and the beam searching can use exactly these fields to suggest charts.

\section{Training and Evaluation Details}
\label{app:exp_setups}
In this section we elaborates detailed setups of experiments in \refsec{sec:exp} for Table2Charts and other baseline methods, including DeepEye~\cite{luo2018deepeye}, Data2Vis~\cite{dibia2019data2vis}, DracoLearn~\cite{moritz2019formalizing} and VizML~\cite{hu2019vizml}.

\subsection{Training Details}

\subsubsection{Training Table2Charts}
\label{sec:t2c-details}

\begin{table}[t]
    \centering
    \setlength{\abovecaptionskip}{3pt}
    \caption{Hyper-parameters of CopyNet Models Sizes.}
    \label{tab:model-size}
    \resizebox{\columnwidth}{!}{
    \begin{tabular}{*{8}{c}}
    \toprule
    Model&\multicolumn{2}{c}{Layers}&\multicolumn{2}{c}{Input Dim}& \multicolumn{2}{c}{Hidden Dim} & Total\\
    \cline{2-7} 
         Size&Encoder&Decoder&Encoder&Decoder&Encoder&Decoder&Parameters \\
    \midrule
    Small & 2 & 1 & 192 & 192 & 128 & 128 & \textasciitilde0.8M\\
    Medium & 2 & 1 & 320 & 256 & 192 & 192 & \textasciitilde1.8M\\
    Large & 4 & 1 & 384 & 512 & 224 & 256 & \textasciitilde4.9M\\
    \bottomrule
    \end{tabular}}
\end{table}

The training process of Table2Charts consists of 30 epochs of teacher forcing followed by 5 epochs of search sampling. First, hyper-parameters of the DQN model are selected by conducting a series of preliminary experiments. For semantic embedding (shown in \reffig{fig:InputEmbed}), two pre-trained NLP embedding models are considered: FastText~\cite{bojanowski2017enriching} with embedding size of 50 and vocabulary size of 200000, and BERT~\cite{Devlin:2018:BERT} with embedding size of 768 and vocabulary (subwords) size of 30522. Besides, we consider three different DQN sizes, with different hidden state dimensions and different number of encoder layers (see \reftab{tab:model-size}).

To choose embedding model and DQN size, we compare their six possible combinations after the teacher forcing training stage on multi-type task.
Results show that compared to FastText, BERT increases R@1 for about 2\% (from 13.07\% to 14.97\%), but doubles the training time. Similarly, R@1 of ``small", ``medium" and ``large" models are 13.07\%, 13.30\% and 15.37\% respectively. The ``large" model gains about 2\% in recalls while the number of parameters is $2.5\times$ ``medium" or $6.1\times$ ``small" model. To make a trade-off between performance and training costs, we use the FastText embedding and the "medium" model size in all experiments in \refsec{sec:exp}.

The beam searching hyper-parameters ($BeamSize$, $ExpandLimit$) are fixed to (4, 100). For neural network tuning, we use Adam optimizer with ($learning\_rate$, ${\beta}_1$, ${\beta}_2$, $\epsilon$, $weight\_decay$) set to ($1e4$, $0.9$, $0.999$, $1e-8$, $0.01$). 
Due to GPU memory limitations, for all experiments the batch size is set to 512. During the back propagation at each step, gradients from each process are averaged.


\subsubsection{Training Data2Vis}
To train and evaluate Data2Vis, we transform Excel and Plotly data to JSON strings. Following data preparation code from Data2Vis, table input is a JSON dictionary containing key-value pairs of field keys to one randomly sampled row, and chart output is a JSON dictionary in a simplified Vega-lite format. For each (table, chart) pair, two samples are generated by sampling two rows from the table. In total, there are 180383 training samples.

Data2Vis uses a character-level seq2seq model with strings as input and output. We set its encoder to be a 1-layer bidirectional LSTM with hidden dimension 256, and decoder to be 2-layer LSTM with hidden dimension 128. These choices make sure the size of the model (\textasciitilde1.94M) is comparable to the that of Table2Charts (\textasciitilde1.8M).


Following the training configurations of Data2Vis, Adam optimizer is used again. According to our Excel corpus, 98\% of the source table JSON strings have fewer than 471 characters, while 99\% target chart JSON strings have fewer than 130 characters. Therefore, the maximum source length and target length are set to 500 and 130. The vocabulary sizes of source and target are 98 and 42. The model is trained for nearly 30000 steps, with a batch size of 16.

\subsubsection{Training VizML}
As mentioned in \refsec{sec:baselines}, VizML focuses on design choices and does not provide models for data queries. We re-train VizML models on its \textit{Mark Type} task (corresponding to chart type) and \textit{Is on X-axis or Y-axis} task (corresponding to field mapping), and change its \textit{Mark Type} task from 2, 3, and 6 classification to 4 classification (including line, scatter, bar, pie chart).
These models make predictions for one field at a time. So only the labelled fields (selected in user-created Excel charts) are kept for training.
Field feature extraction process and model hyper-parameters are identical with the original VizML paper and source code.

\subsection{Evaluation Details}
\label{app:related-work}

\subsubsection{Calculating Recall at Top-k}
In this section, we elaborate how recall numbers are calculated on data queries, design choices and overall chart recommendation tasks in \refsec{sec:MTL-eval} and \refsec{sec:shared-eval}.

On data queries, field selection is compared with ground truth.
Given a table, if the chosen field set of any top-$k$ recommended chart matches that of any user-created chart of the table, this table is considered to be successfully recalled \wrt data queries. Thus, recall at top-$k$ of data queries is calculated as
$R@k = \frac{\text{\#(Tables successfully recalled)}}{\text{\#(Tables)}}$.

On design choices, we consider chart type, field mapping (\ie, map the selected fields onto x-axis and y-axis of a chart), and grouping operation (\ie, whether stacked or clustered, only for bar chart).
Given a table and one set of its fields, if any top-$k$ recommended chart adopts the field set and matches a user-created chart of the table, then the field set is considered to be successfully recalled \wrt design choices.
Thus, the recall on design choices is calculated as 
$R@k = \frac{\text{\#(User-created field sets successfully recalled)}}{\text{\#(User-created field sets)}}$.

On overall evaluation, both data queries and design choices are considered. The equation for overall evaluation is the same as that of data queries, while the ``table successfully recalled'' here means that any top-$k$ recommended chart complete matches any user-created chart of the table. 

\subsubsection{Comparing with Baselines}
For fair comparisons among chart reco systems, more evaluation details (in addition to those described in \refsec{sec:baselines}) need to be properly handled.
In \textbf{DeepEye}, bar grouping operations are not considered, and several data transformations (\eg, dimension breakdown and measure binning) are recommended. To make sure the recommended charts from DeepEye matches the definitions in our corpora, during evaluation we ignore the grouping operations in ground truth and drop the charts with breakdown and binning operations.
In \textbf{Data2Vis}, searching beam size is set to 15 for data query and overall chart recommendation, and set to 30 for design choices. 
In \textbf{DracoLearn}, same as DeepEye, its weights for soft constraints are not re-trained using our Excel training set.
In \textbf{VizML}, we only evaluate it on design choices. Given a field set, if \textit{Mark Type} and \textit{Is on X-axis or Y-axis} predictions of all fields in the set match any of user-created chart, then the field set is considered to be successfully recalled \wrt design choices. Only R@1 is calculated for VizML because only one result is available. 
In \textbf{Table2Charts}, $BeamSize$ and $ExpandLimit$ are the same as in search sampling training (see \refsec{sec:t2c-details}).

\end{appendices}

\end{document}